\documentclass[acmsmall,screen,table,preprint,nonacm]{acmart}
\usepackage{natbib}

\usepackage{booktabs}
\usepackage{subcaption}
\usepackage{siunitx}

\usepackage[linesnumbered,ruled,vlined]{algorithm2e}
\usepackage[utf8]{inputenc}
\usepackage{amsmath}
\usepackage{stmaryrd}

\usepackage{soul}
\usepackage{empheq}
\usepackage[breakable]{tcolorbox}
\usepackage{multirow}
\usepackage{wrapfig}
\usepackage{fancyvrb}
\usepackage{listings}
\usepackage{thmtools,thm-restate}
\usepackage{hhline}
\usepackage{setspace}
\usepackage[normalem]{ulem}
\usepackage{multicol}
\usepackage{graphics}
\usepackage{tcolorbox}
\tcbuselibrary{listings, breakable}

\SetCommentSty{mycommfont}

\newcommand{\bigex}{%
\mathop{\lower0.75ex\hbox{%
   \scalebox{1.7}{\ensuremath{\exists}}}}\limits}

\xspaceaddexceptions{[]\{\}}
\newcommand{\rone}{(\emph{i})\xspace}
\newcommand{\rtwo}{(\emph{ii})\xspace}
\newcommand{\rthree}{(\emph{iii})\xspace}
\newcommand{\rfour}{(\emph{iv})\xspace}
\newcommand{\rfive}{(\emph{v})\xspace}

\let\oldnl\nl
\newcommand{\nonl}{\renewcommand{\nl}{\let\nl\oldnl}}

\newcommand{\Omit}[1]{}

\newcommand{\name}{\texttt{Germinator}\xspace}

\definecolor{dred}{rgb}{0.6,0.0,0.0}
\definecolor{dblue}{rgb}{0.0,0.0,0.5}
\definecolor{dgreen}{rgb}{0.0,0.5,0.0}
\definecolor{gray}{rgb}{0.5,0.5,0.5}

\lstdefinelanguage{MLIR}{
    keywords={func, return},
    keywordstyle=\color{dblue}\bfseries,
    morekeywords={[2]KnownBits, bool},
    keywordstyle={[2]\color{dgreen}\bfseries},
    morekeywords={[3]negate, countLeadingZero, setHighBits, makeKnownBits, unsignedLessThan, ite, meet},
    keywordstyle={[3]\color{dred}},
    sensitive=true,
    comment=[l]{//},
    commentstyle=\color{gray},
    basicstyle=\ttfamily\footnotesize,
    escapeinside=``,
    tabsize=3,
    numbers=none,
    xleftmargin=0em
}

\lstdefinelanguage{llvm}{
    morekeywords={[1]if, return},
    keywordstyle={[1]\color{dblue}\bfseries},
    morekeywords={[2]KnownBits, APInt, unsigned, bool},
    keywordstyle={[2]\color{dgreen}\bfseries},
    morekeywords={[3]countLeadingZero, countTrailingZero, setHighBits, setLowBits, isConstant, getConstant, isPowerOf2, max},
    keywordstyle={[3]\color{dred}},
    sensitive=true,
    comment=[l]{//},
    commentstyle=\color{gray},
    basicstyle=\ttfamily\footnotesize,
    escapeinside=``,
    tabsize=3,
    numbers=none,
    xleftmargin=0em
}

\lstdefinelanguage{TableGen}{
    keywords={def, class, let, in, ins, outs},
    keywordstyle=\color{dblue}\bfseries,
    sensitive=true,
    comment=[l]{//},
    commentstyle=\color{gray},
    basicstyle=\ttfamily\footnotesize,
    escapeinside={(*@}{@*)}
}

\lstdefinelanguage{Lark}{
    keywords={rule, import, token, declare, ignore, \%override, \%import, \%declare, \%ignore},
    keywordstyle=\color{dblue}\bfseries,
    keywordstyle={[2]\color{dgreen}\bfseries},
    sensitive=true,
    comment=[l]{\#},
    commentstyle=\color{gray},
    basicstyle=\ttfamily\footnotesize,
    escapeinside={(*@}{@*)},
    numbers=none,
    xleftmargin=0em,
    alsoletter={:,|,?,*,+},
    literate=
        {:}{{\color{dblue}{:}}}1
        {|}{{\color{dblue}{|}}}1
        {?}{{\color{dblue}{?}}}1
        {*}{{\color{dblue}{*}}}1
        {+}{{\color{dblue}{+}}}1
        {[}{{\color{dblue}{[}}}1
        {]}{{\color{dblue}{]}}}1
        {(}{{\color{dblue}{(}}}1
        {)}{{\color{dblue}{)}}}1
}

\definecolor{magenta}{RGB}{204, 0, 153}

\usepackage{tikz}
\usetikzlibrary{shapes.geometric, arrows.meta, positioning, fit, backgrounds, shadows, decorations.pathreplacing}
\usepackage{xcolor}
\usetikzlibrary{shapes,snakes}
\usetikzlibrary{arrows.meta}
\usetikzlibrary{positioning}
\usetikzlibrary{shapes.geometric}

\definecolor{darkcolor}{HTML}{efefef}
\definecolor{litecolor}{HTML}{c4eded} 

\definecolor{primary}{RGB}{70,130,180}      
\definecolor{secondary}{RGB}{220,100,100}   
\definecolor{tertiary}{RGB}{100,180,100}    
\definecolor{bg1}{RGB}{240,248,255}         
\definecolor{bg2}{RGB}{255,250,240}         
\definecolor{bg3}{RGB}{240,255,240}         

\tikzstyle{process} = [rectangle, rounded corners, minimum width=2.5cm, minimum height=0.8cm, text centered, draw=black, fill=primary!20, font=\small\sffamily]
\tikzstyle{data} = [rectangle, minimum width=2.2cm, minimum height=0.7cm, text centered, draw=black, fill=white, font=\small\sffamily]
\tikzstyle{decision} = [diamond, aspect=2, minimum width=2cm, minimum height=0.8cm, text centered, draw=black, fill=secondary!20, font=\small\sffamily]
\tikzstyle{component} = [rectangle, rounded corners=2mm, draw=black, thick, inner sep=8pt]
\tikzstyle{arrow} = [-{Stealth[length=2mm]}, thick]
\tikzstyle{bigarrow} = [-{Stealth[length=3mm]}, very thick, line width=1.2pt]

\usepackage[framemethod=tikz]{mdframed}
\usepackage{pgfplots}

\usepackage[nosort]{cleveref}

\usepackage{csquotes}
\usepackage{framed}
\usepackage{arydshln}
\usepackage{afterpage}
\usepackage{enumitem}

\newtheorem{example}{Example}[section]
\newtheorem{definition}{Definition}[section]
\setlist[itemize]{align=parleft,left=0pt..1em, topsep=2pt}
\setlist[description]{topsep=2pt}

\begin{document}

\title{Bootstrapping Fuzzers for Compilers of Low-Resource Language Dialects Using Language Models} 

\author{Sairam Vaidya}
\orcid{0009-0000-0609-6507}
\affiliation{%
  \institution{University of California San Diego}
  \country{USA}
}
\email{smahadevaganapathy@ucsd.ed}

\author{Marcel Böhme}
\orcid{0000-0002-4470-1824}
\affiliation{%
  \institution{Max Planck Institute for Security and Privacy}
  \country{Germany}
}
\email{marcel.boehme@mpi-sp.org}

\author{Loris D’Antoni}
\orcid{0000-0001-9625-4037}
\affiliation{%
  \institution{University of California San Diego}
  \country{USA}
}
\email{ldantoni@ucsd.edu}

\begin{abstract}
Modern extensible compiler frameworks---such as MLIR—enable rapid creation of domain-specific language dialects. This flexibility, however, makes correctness harder to ensure 
as the same extensibility that accelerates development also complicates maintaining the testing infrastructure. 
Extensible languages require automated test generation that is both \textit{dialect-agnostic} (works across dialects without manual adaptation) and \textit{dialect-effective} (targets dialect-specific features to find bugs).
Existing approaches typically sacrifice one of these goals by either requiring manually constructed seed corpora for each dialect, or by failing to be effective.

We present a dialect-agnostic and dialect-effective grammar-based and coverage-guided fuzzing approach for extensible compilers that combines two key insights from existing work: 
\rone the grammars of dialects, which already encode the structural and type constraints, can often be extracted automatically from the dialect specification; and 
\rtwo these grammars can be used in combination with pre-trained large language models to automatically generate representative and diverse seed inputs from the full dialect space without requiring any manual input or training data.
These seeds can then be used to bootstrap coverage-guided fuzzers.

We built this approach into a tool, \name.
When evaluated on six MLIR projects spanning 91 dialects, \name generated seeds improve line coverage by 10-120\% over grammar-based baselines. We compare against grammar-based baselines because they are the only class of existing automatic seed generators that can be applied uniformly across MLIR’s heterogeneous dialect ecosystem. 
\name discovers 88 previously unknown bugs (40 confirmed), including 23 in dialects with no prior automated test generators, demonstrating effective and controllable testing of low-resource dialects at scale. \name is available at \url{https://github.com/large-loris-models/germinator}.
\end{abstract}


\maketitle

\section{Introduction} 
\label{se:introduction}

Modern extensible compiler frameworks, such as the Multi-Level Intermediate Representation (MLIR)~\cite{mlir}, allow developers to define custom dialects at the intermediate representation (IR) layer---the ``middle'' of a compiler between source code and machine code. 
Domain experts can prototype new compilation pipelines~\cite{torch-mlir}, integrate novel hardware targets~\cite{circt}, or express domain-specific optimizations~\cite{iree, triton} without reimplementing entire compilers.
This extensibility allows frontends to focus on designing language abstractions for a specific domain without reimplementing low-level compilation infrastructure, thus leveraging the compiler's existing optimizations and code generation~\cite{onnx-mlir, dsp-mlir, quantum-mlir}. 

This extensibility introduces a critical bottleneck: new \textit{dialects} often ship with limited and incomplete test suites, leaving custom dialect features largely untested. 
Compilers are foundational infrastructure, and even small miscompilations can silently propagate correctness failures into production systems~\cite{chengian2016, regehr2011}. 
To avoid correctness bugs, fully verifying the compiler would be ideal, but in practice testing remains the primary method for validating compiler correctness at scale~\cite{xiao2019, chen2020}, and this paper therefore focuses on improving testing for extensible compilers.
Specifically, we focus on fuzzing compilers~\cite{mlir, MLIRSmith, SynthFuzz, suo2025desildetectingsilentbugs, FLEX}.

When trying to build a compiler fuzzer, we typically need to bootsrap the tool with some ``representative'' programs in the language and then randomly mutate (usually following some information about the language) programs to trigger different compiler behaviors. 
Language dialects are by definition \emph{low-resource languages} and lack large program corpora needed to bootsrap the search---i.e., it is hard for the fuzzer to sample ``typical'' programs in each dialect. 
Unlike mature languages with decades of testing infrastructure (e.g., C/C++~\cite{regehr2011, yarpgen}), emerging dialects may have only a handful of example programs and no dedicated ``automatic program generators''. 
Hence, subtle bugs can persist for months, discovered only in production~\cite{Marcozzi_2019}. The core challenge is thus: \emph{how can we automatically test (specifically fuzz) rapidly evolving compiler dialects without incurring substantial manual effort or requiring large training corpora?}


We argue that a good compiler-fuzzer for extensible languages should be:
\begin{itemize}
    \item 
\textit{Dialect-agnostic}---the fuzzer works across different dialects without manual adaptation;
\item
\textit{Dialect-effective}---the fuzzer achieves high coverage and can target specific dialect features that uncover dialect-specific bugs.
\end{itemize}
Table~\ref{tab:properties} summarizes how existing compiler fuzzers require manual input or are dialect-ineffective.

\begin{table}[t]
\centering
\caption{
Comparison of fuzzing approaches for extensible languages. \textit{Dialect-agnostic}: works across dialects without manual adaptation per dialect. \textit{Dialect-effective}: achieves high coverage and can target dialect-specific features to uncover bugs. Note that $\star$ indicates tools that are theoretically dialect-agnostic but \textit{corpus-dependent}.}
\label{tab:properties}
\small
\begin{tabular}{lcc}
\toprule
\textbf{Approach} & \textbf{Dialect-agnostic} & \textbf{Dialect-effective} \\
\midrule
Grammarinator~\cite{grammarinator} & $\star$& $\times$ \\
SynthFuzz~\cite{SynthFuzz} & $\star$ & \checkmark  \\
FLEX~\cite{FLEX} & $\times$ & \checkmark \\
MLIRSmith~\cite{MLIRSmith} & $\times$ & \checkmark \\
MLIRod~\cite{MLIRod} & $\times$ & \checkmark \\
\midrule
\textbf{\name} & \checkmark & \checkmark \\
\bottomrule
\end{tabular}
\end{table}

The main insight for our approach is that effective compiler fuzzing for low-resource dialects fundamentally hinges on starting from a small set of seed programs whose distribution approximates that of the dialect under test, and then using coverage-guided fuzzers---i.e., a tool that mutates programs to exercise different execution paths---to randomly mutate such seeds to trigger rare events.
That is, our approach reframes the compiler fuzzing problem as one of automatically generating high-quality, diverse seed programs---representative examples in a given dialect---rather than relying on manually constructed corpora as in prior work~\cite{MLIRod, SynthFuzz, FLEX}.

Following the above insight, our key innovation is to demonstrate that large language models (LMs) can serve as effective samplers even for low-resource dialects, provided that the language models are constrained to produce programs that adhere to the dialect's syntax.
This syntactic restriction can be enforced by providing the model with a context-free grammar for the dialect's syntax and applying grammar-constrained sampling techniques~\cite{dong2023codepad, geng-etal-2023-grammar, scholak2021picard, poesia2022synchrome, park2025flexible, parkGad, syncode, mcmceasy, GuidanceAI2023, awrs} to ensure that all generated programs are syntactically valid.
In essence, our method combines the strengths of grammar-based fuzzers (which only sample syntactically valid programs) and coverage-guided fuzzing (which take into account coverage feedback to make sample effectively).

We implement our approach in a tool, \name, that is both \textit{dialect-agnostic} and \textit{dialect-effective}.
Given a target dialect for the LLVM compiler framework, \name automatically extracts from the dialect's TableGen definitions a context-free grammar encoding the dialect's syntactic constraints. 
A TableGen definition is a declarative specification in LLVM's domain-specific language that defines operations, their signatures, type constraints, and assembly syntax.
Using the extracted grammar, \name employs state-of-the-art grammar-constrained sampling algorithms~\cite{GuidanceAI2023} over pre-trained language models to generate diverse and syntactically valid seeds---without requiring additional training, fine-tuning, or large manually curated corpora.
The generated seeds are then passed to an existing coverage-guided fuzzer.

We evaluate \name on six MLIR-based projects spanning 91 dialects and 1338 operations: base MLIR~\cite{mlir}, Torch-MLIR~\cite{torch-mlir}, IREE~\cite{iree}, CIRCT~\cite{circt}, Triton~\cite{triton}, and HEIR~\cite{heir}. 
\name discovered 88 previously unknown bugs (40 confirmed), including 23 in dialects with no prior automated test generators. Compared to the best-performing baselines, \name improves line coverage by 10--120\%  (see \Cref{sec:evaluation}).


\textbf{Contributions.} This paper makes the following contributions:
\begin{itemize}
    \item We formulate the challenge of designing compiler fuzzers that generalize across dialects---\textit{dialect-agnostic}---while remaining effective within each specific dialect---\textit{dialect-effective} (\Cref{sec:overview}).

    \item We provide a probabilistic formulation of compiler fuzzing that shows that coverage-guided fuzzers are \textit{dialect-effective}, provided that they begin from representative seeds. We thus reduce \textit{dialect-agnostic} fuzzing to the automatically sampling representative dialect seeds (\Cref{sec:approach}).

    \item We show how to build effective seed samplers by combining large language models with grammar-constrained decoding algorithms that force such models to only produce programs accepted by a given context-free grammar describing the dialect's syntax (\Cref{sec:sampling}).

    \item We implement our approach in \name, a \textit{dialect-agnostic} and \textit{dialect-effective} tool that automatically extracts grammars from TableGen definitions and generates high-quality seeds using grammar-constrained sampling with language models (\Cref{sec:implementation}).
    
    \item Our evaluation demonstrates that \name is 
    \textit{dialect-agnostic}---extracting grammars for 88.9\% of operations across all 91 MLIR dialects---and 
    \textit{dialect-effective}---it improved line coverage by 10-120\% when compared to best-performing baselines and discovered 88 bugs (40 confirmed, 14 repaired) across six projects, including 23 bugs in dialects with no existing generators.
\end{itemize}

\section{Overview: Testing an MLIR Dialect with Our Approach}
\label{sec:overview}


\begin{wrapfigure}{r}{0.5\textwidth}
    \vspace{-0.6em}
    \centering
\begin{lstlisting}[language=MLIR, basicstyle=\small\ttfamily,
  escapeinside={(*@}{@*)}, frame=single]
func.func @test() {
  %0 = emitc.literal "0" : index
  %1 = emitc.literal "10" : index  
  %2 = emitc.literal "1" : index
  (*@\colorbox{red!20}{emitc.for \%iv = \%0 to \%1 step \%2 \{}@*)
    (*@\colorbox{red!20}{// Region body misses block arg for \%iv}@*)
    emitc.yield
  (*@\colorbox{red!20}{\}}@*)
  return
}
\end{lstlisting}
\caption{A valid \texttt{emitc.for} loop missing its induction variable. The verifier assumes this variable always exists and crashes when trying to access it.}
    \vspace{-0.6em}
\label{fig:emitc-example}
\end{wrapfigure}

We illustrate our approach through a concrete example: automatically testing the \texttt{EmitC} dialect in MLIR, a low-resource dialect that translates MLIR constructs into C/C++ code~\cite{emitc-docs, mlir}.
The \texttt{EmitC} dialect is used by developers who need to lower MLIR-based IRs into portable C/C++ backends---for example, to integrate MLIR pipelines into embedded, safety-critical, or cross-platform toolchains where LLVM is unavailable.
%
With only 47 test cases in its official repository~\cite{emitc-tests}, \texttt{EmitC} epitomizes the testing bottleneck faced by small or emerging dialects whose rapid evolution outpaces manual test creation: in just the past year, \texttt{EmitC} has seen over one hundred pull requests merged with multiple API changes---yet only \textit{a few dozen new tests}---highlighting the difficulty of maintaining test coverage as dialect semantics shift.

In the \texttt{EmitC} dialect, control-flow constructs mirror the structure of C code. 
An \texttt{emitc.for} operation  (see \Cref{fig:emitc-example}) corresponds to a C-style \texttt{for} loop, where the body represents the loop body and the loop bounds specify the initialization, condition, and increment expressions.
In typical usage, the loop body receives an induction variable argument that represents the loop counter.
However, in the program in \Cref{fig:emitc-example}, the body takes no arguments---yielding an unusual but valid loop form: it iterates over a fixed range without exposing the induction variable to its body.

This example exposes a real bug that \name automatically discovered in the \texttt{EmitC} compiler (link omitted for anonymity).
The program is syntactically valid: the \texttt{emitc.for} operation accepts three operands (two bounds and a step) and a region. However, the program triggers an assertion failure because the verifier's \texttt{getInductionVar()} method assumes a block argument exists without checking.
This reveals a spec-implementation gap: the TableGen specification permits loops without block arguments, but the verifier implementation assumes one always exists. Such mismatches arise when verification code makes implicit assumptions not enforced by the formal specification.

\subsection{Why Existing Approaches Fail}
\label{sec:why-existing-fail}

We explain why existing fuzzing approaches (\Cref{tab:properties}) do not uncover this bug and similar ones.

\textit{Grammar-based fuzzers}, e.g., Grammarinator~\cite{grammarinator} and SynthFuzz~\cite{SynthFuzz}, generate programs from context-free grammars, either manually defined or mined. In Grammarinator, initial seed programs are retained and mutated, biasing generation toward known structures. While this approach avoids syntactically invalid programs, semantic constraints---such as type correctness and valid region usage---are often violated. Consequently, constructs like an \texttt{emitc.for} loop with the precise operand types and region structure needed to trigger our bug are rarely produced.
SynthFuzz extends grammar-based fuzzing with context-dependent mutations learned from seeds, increasing the likelihood of valid programs. 
However, in low-resource dialects like \texttt{EmitC} only a few seeds contain relevant constructs (e.g., 3 \texttt{emitc.for} loops), and the fuzzer is unlikely to produce the exact combination required to reach the bug.

\textit{Hand-engineered fuzzers}, e.g., MLIRSmith~\cite{MLIRSmith} and MLIRod~\cite{MLIRod}, ensure high syntactic and semantic validity through dialect-specific type systems and dependency models. 
However, neither tool supports the \texttt{EmitC} dialect.
To generate our running example one would need to \textit{manually add} \texttt{EmitC}-specific rules---an effort that is prohibitive for low-resource or rapidly evolving dialects.

\textit{Learning-based generators}, like FLEX~\cite{FLEX}, learn patterns from large corpora, producing syntactically valid programs that satisfy semantic constraints. Their effectiveness depends on hundreds or thousands of seeds; in low-resource dialects like \texttt{EmitC}, these tools overfit and cannot generalize to rare constructs such as loops without induction variables.

In summary, grammar-based fuzzers ignore semantic constraints, hand-engineered fuzzers require prohibitive per-dialect effort, and learning-based generators fail in low-resource settings.



\subsection{Why Our Approach Succeeds}
\label{sec:why-we-succeed}

\begin{figure}[t]
\centering
\begin{lstlisting}[language=TableGen, basicstyle=\footnotesize\ttfamily, frame=single,
  escapeinside={(*@}{@*)},
  moredelim={**[is][\color{blue}]{∫}{∫}},
  moredelim={**[is][\color{magenta}]{®}{®}},
  caption={(a) \texttt{EmitC} TableGen specifications}]
def EmitC_LiteralOp : EmitC_Op<"literal", [Pure]> {
  // explicit format -> direct extraction 
  ∫let assemblyFormat = "$value attr-dict `:` type($result)";∫ ...
}
def EmitC_ForOp : EmitC_Op<"for"> {
  let arguments = (ins ∫IntegerIndexOrOpaqueType:$lowerBound,∫ ...);
  ®let hasCustomAssemblyFormat = 1;® ... // custom printer -> needs inference
}
def EmitC_YieldOp : EmitC_Op<"yield", [Terminator]> {
  // explicit format -> direct extraction 
  ∫let assemblyFormat = "attr-dict ($result^ `:` type($result))?";∫ ...
}
\end{lstlisting}

\vspace{0.1cm}

\begin{lstlisting}[language=Lark, basicstyle=\footnotesize\ttfamily, frame=single,
  escapeinside={(*@}{@*)},
  moredelim={**[is][\color{blue}]{∫}{∫}},
  moredelim={**[is][\color{magenta}]{ø}{ø}},
  caption={(b) Extracted grammar rules}]
# Directly extracted from assemblyFormat
∫emitc_literal∫: ∫op_result_list∫? ∫"emitc.literal" ... attribute_dict∫? ":" ∫type∫
∫integer_index_or_opaque∫: ∫"index"∫ | ∫"i1"∫ | ... | ∫"i128"∫
∫emitc_yield∫: ∫"emitc.yield"∫ ∫attribute_dict∫? (∫ssa_use∫ ∫"∫∫:∫∫"∫ ∫type∫)?

# Inferred from custom format + LM
øemitc_forø: ø"emitc.for"ø øSSA_ID "=" ssa_use "to" ssa_use "step"ø ... ø?ø øregionø

\end{lstlisting}

\caption{
Grammar extraction for the \texttt{EmitC} dialect. (a) Snippet of TableGen operation definitions: directly extracted assembly formats are highlighted in \textcolor{blue}{blue} and custom C++ formats requiring inference are highlighted in \textcolor{magenta}{magenta}. (b) Grammar rules extracted from TableGen enforcing syntactic constraints, while semantic constraints---such as operand compatibility---are handled implicitly by the language model.
}
\label{fig:grammar-extraction}
\end{figure}

\Cref{sec:why-existing-fail} makes it clear that the core challenge in fuzzing compilers is to generate programs that are simultaneously \textit{diverse}---covering unexplored regions of the dialect's design space---and \textit{valid}---respecting both syntactic and semantic constraints---without relying on extensive human-written corpora or manual engineering.
\name succeeds where existing fuzzers fail by addressing both dimensions: using pre-trained language models to achieve program \emph{diversity} and grammar-constrained sampling to ensure syntactic and semantic \emph{validity}---effectively combining the best ideas from machine learning and classical compiler testing.


\subsubsection{Mining Grammars to Enforce Validity}
First, \name automatically extracts a context-free grammar from the dialect's \texttt{TableGen} specification (see \Cref{fig:grammar-extraction}). 
\texttt{TableGen} files declaratively encode the syntax of operations, types, and attributes. 
For the \texttt{TableGen} file in \Cref{fig:grammar-extraction}(a), \name identifies assembly formats and type vocabularies for which grammar rules can be directly extracted (highlighted in \textcolor{blue}{blue}) as well as custom formats for which extracting a grammar rule requires more complex inference using language models (highlighted in \textcolor{magenta}{magenta}). 
These elements are converted into grammar rules (\Cref{fig:grammar-extraction}(b)) that enforce syntactic constraints, ensuring generated programs adhere to the dialect's structural rules. 
Patterns not fully captured by the grammar---such as optional attributes, complex control-flow combinations, or region-specific constraints---are handled in the next step via language model sampling, which takes advantage of learned structural priors.

\subsubsection{Constrained Sampling with Language Models for Diverse and Valid Seed Generation}
Second, \name takes advantage of the broad knowledge embedded in pre-trained  language models (LMs) to generate diverse programs.
Because pre-trained models have already learned general MLIR patterns from large corpora, they provide a strong starting point even for previously unseen dialects.
However, because the dialects are out of the model's learned distribution, the model alone cannot enforce dialect-specific constraints and will mostly produce samples that are invalid in the dialect (\Cref{fig:gcd-comparison})---i.e., language models suffer the same limitations as existing approaches.

\name uses recent advances in \emph{grammar-constrained sampling}~\cite{geng-etal-2023-grammar, dong2023codepad, poesia2022synchrome, scholak2021picard, parkGad, park2025flexible, GuidanceAI2023} to guide the language model to only produce programs that are valid with respect to the grammar extracted in the first step, effectively combining the strengths of grammar-based (through the grammar) and learning-based approaches (the universal knowledge learned by the language model).
The grammar enforces syntactic validity wherever explicit rules exist, while the LM introduces diversity and statistical patterns not captured by the grammar, making \name a universal generator for low-resource dialects.

\begin{figure}[t]
\centering
\begin{lstlisting}[basicstyle=\footnotesize\ttfamily, frame=single, numbers=none]
Prompt: Generate a short test case directly in MLIR in assembly format 
        for the EmitC dialect that uses an emitc.for loop with literals as bounds.
\end{lstlisting}

\vspace{0.2cm}
\begin{minipage}[t]{0.48\textwidth}
\begin{lstlisting}[language=MLIR, basicstyle=\scriptsize\ttfamily, frame=single,
  title={(a) Unconstrained},
  escapeinside={(*@}{@*)},
  backgroundcolor=\color{red!5}]
module @test {
  (*@\colorbox{red!20}{!i32 3 = arith.constant 3}@*)
  (*@\colorbox{red!20}{!i32 5 = arith.constant 5}@*)
  %i = (*@\colorbox{red!20}{emitc.for \%i in [\%3, \%4]}@*)
    : (i32) to (i32) step (i32 1) {
    (*@\colorbox{red!20}{!i32 10 = arith.constant 10}@*)
    (*@\colorbox{red!20}{emitc.print \%10 : (i32)}@*)
  }
  (*@\colorbox{red!20}{return \%i : (i32)}@*)
}
\end{lstlisting}
\end{minipage}
\hfill
\begin{minipage}[t]{0.48\textwidth}
\begin{lstlisting}[language=MLIR, basicstyle=\scriptsize\ttfamily, frame=single,
  title={(b) Grammar-Constrained},
  escapeinside={(*@}{@*)},
  backgroundcolor=\color{green!5}]
func.func @emitc_example() -> i32 {
  (*@\colorbox{green!20}{\%i = emitc.literal "5" : i32}@*)
  (*@\colorbox{green!20}{\%j = emitc.literal "10" : i32}@*)
  (*@\colorbox{green!20}{\%k = emitc.literal "2" : i32}@*)
  (*@\colorbox{green!20}{emitc.for \%iv = \%i to \%j}@*)
    (*@\colorbox{green!20}{step \%k \{}@*)
    (*@\colorbox{green!20}{emitc.yield}@*)
  (*@\colorbox{green!20}{\}}@*)
  return %i : i32
}
\end{lstlisting}
\end{minipage}

\caption{Grammar-constrained sampling ensures syntactic validity. Given the same prompt (the specific prompt is not one used by \name and is for illustrative purposes), (a) unconstrained LLM generation produces multiple syntax errors (\textcolor{red}{highlighted in red}): incorrect SSA value declarations (\texttt{!i32 3 = ...}), invalid loop syntax (\texttt{in [\%3, \%4]}), non-existent operations (\texttt{emitc.print}), and invalid module-level return. (b) Grammar-constrained sampling (\textcolor{green!50!black}{highlighted in green}) follows the extracted grammar rules exactly, producing syntactically valid code that respects the EmitC dialect's constraints.}
\label{fig:gcd-comparison}
\end{figure}

\subsubsection{Coverage-guided Fuzzing for Diverse Generation}
Because grammar-constrained LM sampling is computationally (and financially!) expensive, we can typically generate only a few hundred/thousand seeds in reasonable time (minutes) in the second step.

In the third and final step, the hundreds of high-quality seed programs generated by the language model are used as seeds for a coverage-guided fuzzer, which can cheaply produce thousands or millions of additional inputs through mutation and coverage feedback.
Since the seeds already represent meaningful, well-formed compiler programs, a coverage-guided fuzzer can use coverage feedback from the compiler to efficiently explores variations of the existing programs that explore new execution paths and uncovers subtle bugs, such as the induction-variable issue shown earlier in \Cref{fig:emitc-example} that prior approaches missed. 

For the \texttt{EmitC} dialect, \name achieves approximately 75\% higher code coverage than baseline tools and uncovers a previously unknown bug, demonstrating its ability to generate semantically valid programs that exercise rare control-flow and type patterns.

\section{Fuzzing as Distribution Exploration}
\label{sec:approach}


We now formalize the problem tackled in this paper and the intuition from \S\ref{sec:overview} by viewing the task of fuzzing a compiler for a dialect of an extensible language as the problem of efficiently exploring a distribution over programs in that dialect. This perspective clarifies what makes a fuzzer effective, why extensible languages make this setting uniquely challenging, and motivates the design principles underlying our approach.

\subsection{Fuzzing as Sampling from a Program Distribution}

We model an extensible language by starting from a \emph{core language} and defining dialects as extensions of it.
The core of the extensible language $\mathcal{C}$ is described by a grammar $G$, which generates its programs, and a set of validity rules $\varphi$ (e.g., typing or well-formedness checks).  
The set of valid programs in the core language is defined as
\[
L \;=\; \{\, p \mid p \text{ is generated by } G \text{ and } p \text{ satisfies } \varphi \,\}.
\]

A dialect $\mathcal{D}$ defines additional constructs independent of the core language $\mathcal{C}$. It has its own grammar $G_{\mathcal{D}}$ that can generate dialect-specific operations, types, and attributes, along with validity rules $\varphi_{\mathcal{D}}$.
The set of valid dialect programs is
\[
L_{\mathcal{D}} \;=\; \{\, p \mid p \text{ is generated by } G_{\mathcal{D}} \text{ and } p \text{ satisfies } \varphi_{\mathcal{D}} \,\}.
\]

In practice, MLIR programs mix constructs from multiple dialects and the core language by explicitly registering the required dialects in the compiler context. 
Therefore, programs using dialect $\mathcal{D}$ may contain dialect-specific operations mixed with operations in the core language. 
Specifically, dialects are \emph{extensions} in the sense that the $G_{\mathcal{D}}$ contains a superset of the productions of the grammar $G$, and, similarly the semantic constraints are the form $\varphi_{\mathcal{D}} =\varphi \wedge \varphi_{new}$, where the semantic constraints $\varphi_{new}$ are those added by the dialect.



\begin{example}[EmitC composes with core MLIR]
The core language is MLIR, described by grammar $G_{\mathrm{MLIR}}$ and validity rules $\varphi_{\mathrm{MLIR}}$ (for example, productions and typing rules for \texttt{func.func}, \texttt{func.return}, etc.).  
The \texttt{EmitC} dialect extends MLIR by adding dialect-specific productions (for instance, rules for \texttt{emitc.for}, \texttt{emitc.literal}, and \texttt{emitc.yield}) and additional semantic constraints. Concretely, the dialect grammar $G_{\mathrm{EmitC}}$ contains the productions of $G_{\mathrm{MLIR}}$ together with the new \texttt{EmitC} productions, and its validity rules take the form
\(\varphi_{\mathrm{EmitC}} = \varphi_{\mathrm{MLIR}} \wedge \varphi_{\mathrm{EmitC,new}}\),
where $\varphi_{\mathrm{EmitC,new}}$ captures \texttt{EmitC}-specific invariants (e.g., constraints on loop bounds or type relations).
\end{example}

\subsubsection*{Program distributions and coverage.}
Developers do not write programs uniformly at random: some constructs are idiomatic and frequent, while others are rare boundary cases.  
We model this aspect using a (latent) \emph{program distribution}
$P_{\mathcal{D}} : L_{\mathcal{D}} \rightarrow [0,1]$,
where $P_{\mathcal{D}}(p)$ reflects how likely program $p$ is to appear in realistic use of the dialect.

A \textit{gray-box coverage-guided compiler fuzzer} samples programs $p \in L_{\mathcal{D}}$ and executes them to probe the compiler~\cite{even2023grayc, jitfuzz}.  
To quantify how much a program $p$ exercises the compiler, the fuzzer can measure its \emph{coverage}: the parts of the compiler implementation executed when compiling the program $p$.  
Let $\mathcal{E}$ denote the set of coverage events (e.g., basic blocks or branches) we are interested in triggering.  
For a program $p$, let $\mathrm{Cov}(p) \subseteq \mathcal{E}$ be the events it covers, and for a set of programs $S$, let $\mathrm{Cov}(S)=\cup_{p\in S} \mathrm{Cov}(p)$ be the union of the individual programs' coverage.
A fuzzer is effective if it produces a set of samples $S$ that achieves large $|\mathrm{Cov}(S)|$.

In practice, effective compiler fuzzing requires evaluating millions of programs and a fuzzer must generate valid programs at high throughput; even a small slowdown or low validity rate makes large-scale exploration infeasible in practice.
Typically, we are interested in the set of samples $S$ being produced over a finite amount of time---e.g., 24 hours.
Simply ``searching for programs that trigger new events'' is intractable.
Effective fuzzing therefore requires the ability to \emph{sample} from a distribution that \rone resembles $P_{\mathcal{D}}$---to generate realistic programs---and \rtwo upweights programs that exercise rare behaviors---those that $P_{\mathcal{D}}$ assigns low probability to but are crucial for testing.

To formalize this goal, we define a \emph{coverage-oriented target distribution}:
\begin{definition}[Coverage-Weighted Target Distribution]
\label{def:target-distribution}
The \emph{coverage weight} of a program $p$ is
\[
cw(p) \;:=\; \sum_{e \in \mathrm{Cov}(p)} \frac{1}{\Pr_{p \sim P_{\mathcal{D}}}[e \in \mathrm{Cov}(p)]},
\]
and the \emph{coverage-weighted target fuzzing distribution} is
\[
P^\star_{\mathcal{D}}(p) \;\propto\; P_{\mathcal{D}}(p) \cdot cw(p), \qquad p \in L_{\mathcal{D}}.
\]
\end{definition}

Here, $\propto$ means \textit{proportional to}.
Intuitively, $P^\star_{\mathcal{D}}$ preserves the realistic structure and idioms of $P_{\mathcal{D}}$, but amplifies rare behaviors (thus increasing coverage) that are valuable for testing.  
For simplicity, \Cref{def:target-distribution} assumes that every coverage event has non-zero probability under $P_{\mathcal{D}}$.

This distributional perspective provides a unifying lens: fuzzers can be understood by how well they approximate $P^\star_{\mathcal{D}}$.
A \emph{fuzzing strategy} $F$ for an extensible language $\mathcal{C}$ is any procedure that given a dialect $\mathcal{D}$ of the core language generates programs in $L_{\mathcal{D}}$; we denote by $F_{\mathcal{D}}$ the distribution over $L_{\mathcal{D}}$ induced by the strategy's program generation when targeting dialect $\mathcal{D}$.

\begin{definition}[Dialect-Agnostic and Dialect-Effective Fuzzing Strategy]
\label{def:da-de-fuzzing}
A fuzzing strategy $F$ for dialects in core language $\mathcal{C}$ should be:
\begin{itemize}
\item \textbf{Dialect-agnostic.} $F$ requires no manual, dialect-specific engineering for each dialect $\mathcal{D}$: no hand-written generators, encodings, or bespoke type rules crafted by experts.

\item \textbf{Dialect-effective.} For each dialect $\mathcal{D}$, the  distribution $F_{\mathcal{D}}$ induced by $F$ approximates the target distribution $P^\star_{\mathcal{D}}$, yielding inputs that achieve high coverage on the dialect's compiler.
\end{itemize}
\end{definition}

When considering the approaches in \Cref{sec:why-existing-fail}, grammar-based~\cite{grammarinator,SynthFuzz} and learning-based fuzzers~\cite{FLEX} are not dialect-effective: the former generate mostly invalid programs because they ignore semantic constraints, while the latter collapse onto frequent patterns when training data is scarce.
Conversely, hand-engineered fuzzers~\cite{MLIRSmith,MLIRod} can be dialect-effective but are not dialect-agnostic, as they require substantial per-dialect manual engineering of syntax and type rules.

\subsection{Coverage-Guided Fuzzing is Dialect-Effective (With the Right Seeds)}
\label{sec:cgf}


Most practical compiler fuzzers---whether grammar-based, learning-based, or hand-engineered---embed the same underlying mechanism: a \emph{coverage-guided loop} that mutates a seed corpus and retains programs that increase coverage.  
Rather than being a separate class of techniques, coverage guidance is the common amplifier that all these systems rely on.

From the distributional perspective, the coverage-guided loop adjusts the induced sampling distribution $F_{\mathcal{D}}$ toward $P^\star_{\mathcal{D}}$: once a program with high coverage weight $cw(\cdot)$ is discovered, mutation operators generate nearby variants, increasing the probability mass assigned to regions of the program space that trigger rare behaviors. 
Algorithm~\ref{alg:cgf} provides a simplified view of this process.

\begin{algorithm}[t]
\caption{Coverage-Guided Fuzzing Loop}
\label{alg:cgf}
\DontPrintSemicolon
\KwIn{Dialect $\mathcal{D}$; seed generator $\mathtt{gen}$; mutator $\mathsf{Mutate}$; time budget $T$}
\KwOut{Corpus $S \subseteq L_{\mathcal{D}}$ with monotonically increasing $\mathrm{Cov}(S)$}
\BlankLine

$S \gets \mathtt{gen}(\mathcal{D})$ \label{alg:init-seeds} \tcp*[r]{Initial seeds}
\While{time $< T$}{
  $p \sim S$  \label{alg:sample} \tcp*[r]{Sample a program to mutate}  
  $q \gets \mathsf{Mutate}(p)$ \label{alg:mutate} \tcp*[r]{Mutate $p$}  
  \If{$\mathrm{Cov}(q) \nsubseteq \mathrm{Cov}(S)$}{
     $S \gets S \cup \{q\}$ \label{alg:accept}\tcp*[r]{Add $q$ to $S$ if it increases coverage}
  }
}
\Return $S$
\end{algorithm}

The effectiveness of this amplification relies on a structural property of compilers for extensible languages: \emph{programs that trigger rare behaviors are typically syntactically close to common programs in the dialect}. That is, reaching new coverage often requires only small edits---e.g., changing an operand, altering a type, or adding a dialect-specific operation. Local mutation is therefore well-suited to refine a seed distribution that already resembles $P_{\mathcal{D}}$.
More formally, for every coverage event $e \in \mathcal{E}$ there exists a program $q \in L_{\mathcal{D}}$ that covers $e$---i.e., $e \in \mathrm{Cov}(p)$---and is syntactically close to a likely program in the dialect distribution---i.e., there exists a program $p \in L_{\mathcal{D}}$ such that $P_{\mathcal{D}}(p)\gg P_{\mathcal{D}}(q)$  and the edit distance $d(p,q)$ is small.

Because mutation operators apply only local edits and are entirely automated (and can even rely on user-provided grammars to avoid syntactic mistakes), the quality of the initial seed set is crucial.
This is one of the key reasons why learning-based fuzzers and hand-engineered fuzzers are effective on the dialects they are tailored to: they can rely on high-quality seeds that approximate the dialect distribution $P_{\mathcal{D}}$.

\begin{tcolorbox}[title=\textbf{Key Observation: Seeds Make or Break Coverage-Guided Fuzzing}, colback=white]
Coverage-guided fuzzing can \emph{amplify} a good seed distribution, but it cannot \emph{repair} a bad one.  
If seeds induce a distribution close to $P_{\mathcal{D}}$, then coverage guidance efficiently steers toward $P^\star_{\mathcal{D}}$, discovering rare high-value behaviors.  
If the seeds lack basic coverage of common dialect constructs, mutation is unlikely to reach the rare behaviors.

\medskip
\textit{Thus, to obtain a dialect-agnostic and dialect-effective strategy, it suffices to produce a small set of seeds whose induced distribution approximates the dialect's program distribution $P_{\mathcal{D}}$.}
\end{tcolorbox}

The above observation reduces the challenge of dialect-agnostic and dialect-effective fuzzing to constructing a \emph{good seed generator}: a sampling mechanism $\mathtt{gen}$ that for any dialect $\mathcal{D}$, without manual engineering, even when producing a small numbers of seeds, induces a distribution that is representative of the dialect's distribution.

\begin{definition}[Good Seed Generator]
\label{def:seed-generator}
A \emph{seed generator} $\mathtt{gen}$ is a sampling algorithm that given a dialect $\mathcal{D}$, can sample seed programs. A \textit{good seed generator} should be:
\begin{itemize}
    \item \textbf{Dialect-agnostic:} requires no manual per-dialect engineering.
    \item \textbf{Representative:} samples seeds approximately proportionally to the dialect's distribution $P_{\mathcal{D}}$.
    \item \textbf{Efficient:} produces hundreds of seeds in minutes.
\end{itemize}
\end{definition}
Efficiency is necessary to avoid delaying the fuzzer's mutation phase. Representativeness is critical because modern fuzzers execute $10^4$–$10^7$ programs per minute; even modest reductions in seed validity or valid-program generation throughput (e.g., from hundreds to tens of valid programs per minute) severely limit the reachable coverage within a fixed fuzzing budget.

\section{Seed Generation via Grammar-Constrained Language Model Sampling}
\label{sec:sampling}

Having reduced dialect-effective fuzzing to constructing a seed generator that satisfies the criteria in \Cref{def:seed-generator}, we now present a practical realization that is \emph{dialect-agnostic}, \emph{representative}, and \emph{efficient}. 
Our approach combines pre-trained \emph{language models} (LMs)---which act as dialect-agnostic and distributionally representative samplers---with \emph{grammar-constrained sampling}, which enforces syntactic validity and thus efficiency.

\subsection{Generating Seeds using Language Models}
(Large) Language models (LMs) trained on code corpora learn  programming regularities (e.g., variable binding, SSA structure, type-coherent arithmetic) and host-language idioms (e.g., C/C++/LLVM syntax).  
When an LM has internalized the \emph{core language} $\mathcal{C}$, prompting it with minimal context about a dialect $\mathcal{D}$ can already yield programs that \textit{resemble} valid dialect programs.

Formally, the LM induces a distribution $Q_{\text{LM}} : \Sigma^\ast \rightarrow [0,1]$ over strings, where $\Sigma$ is a set of characters (typically called tokens).  
Typically we are interested in using LMs with input prompts and sample possible continuations of such prompts.
For notation simplicity, in this section we assume a fixed ``prompt'' $\pi$ and the LM only produces outputs for that particular prompt---i.e., $Q_{\text{LM}}(p)$ is actually the probability that $LM$ produces program $p$ on the fixed prompt $\pi$.
When projected onto the set of valid programs $L_{\mathcal{D}}$, its mass
\begin{equation}
\label{eq:llm-lq}
Q_{\text{LM}\mid L_{\mathcal{D}}}(p)=
   \frac{Q_{\text{LM}}(p)\cdot\mathbf{1}[x \in L_{\mathcal{D}}]}{{\sum_{p' \in L_{\mathcal{D}}} Q_{\text{LM}}(p')}}
\end{equation}
tends to correlate with $P_{\mathcal{D}}(p)$ because both distributions are shaped by the same host-language statistics and training data---i.e., the model is trained on data that is representative of how people write code.
Formally, we write that is $Q_{\text{LM}\mid L_{\mathcal{D}}} \approx P_{\mathcal{D}}$.  
In other words, though of course such a statement is not formally provable, the LM approximates the marginal of $P_{\mathcal{D}}$ on shared syntax and ``fills in'' dialect-specific gaps with semantically nearby constructs.

This aspect makes LMs inherently \emph{dialect-agnostic} and, at a high level, \emph{representative}: their samples respect realistic structural priors of the programming ecosystem.  
However, as shown in \Cref{fig:gcd-comparison}, na\"ive sampling from $Q_{\text{LM}\mid L_{\mathcal{D}}}$ is inefficient because an LM is unlikely to sample valid programs in $L_{\mathcal{D}}$ for low-resource dialects and will often produce programs that are not in the grammar $G_{\mathcal{D}}$ or programs that violate semantic constraints $\varphi_{\mathcal{D}}$.


\subsection{Generating Valid Seeds Using Grammar-Constrained Sampling}
To satisfy the \emph{efficiency} requirement of a good seed generator, we need to condition the LM distribution to only (or at least mostly) sample valid programs in $L_{\mathcal{D}}$---i.e., we would like to directly sample from the distribution $Q_{\text{LM}\mid L_{\mathcal{D}}}$ and not generate many invalid programs along the way.

Given a context-free grammar \(G_{\mathcal{D}}\) that specifies the dialect's concrete syntax, we make LM sampling \emph{syntax-aware} by restricting the model to produce only sequences in the language defined by $G_{\mathcal{D}}$. 
This technique is called grammar-constrained sampling~\cite{dong2023codepad, geng-etal-2023-grammar, scholak2021picard, poesia2022synchrome, park2025flexible, parkGad, syncode, GuidanceAI2023}, and intuitively it is implemented by augmenting the LM's generative process with a parser that throws away ungrammatical sequence. 
At each generation step, the model proposes candidate next tokens (i.e., symbols to add to the output being produced) according to its learned distribution, while the grammar's parser determines which of these tokens keep the generate string in the possible language induced by the grammar. 
Invalid continuations---those that would lead to a string outside the grammar's language---are excluded from consideration.  

Grammar-constrained sampling can draw from the following \emph{conditional distribution}:
\begin{equation}
\label{eq:llm-gq}
Q_{\text{LM}\mid G_{\mathcal{D}}}(p)=
   \frac{Q_{\text{LM}}(p)\cdot\mathbf{1}[x \in G_{\mathcal{D}}]}{{\sum_{p' \in G_{\mathcal{D}}} Q_{\text{LM}}(p')}}
\end{equation}
The distribution $Q_{\text{LM}\mid G_{\mathcal{D}}}(p)$ is not the same distribution as $Q_{\text{LM}\mid L_{\mathcal{D}}}(p)$ because it may sample syntactically valid programs that do not satisfy the semantic constraints $\varphi_\mathcal{D}$. 
However, because every program in  $L_{\mathcal{D}}$ is also syntactically valid an in $G_\mathcal{D}$, we have that the two distributions are identical with respect to $L_\mathcal{D}$---i.e., $Q_{\text{LM}\mid L_{\mathcal{D}}}=Q_{\text{LM}\mid G_\mathcal{D} ,L_{\mathcal{D}}}$.
Therefore, grammar-constrained sampling provides a strictly more efficient way to obtain valid programs than na\"ive unconstrained sampling.\footnote{\citet{parkGad} showed that na\"ive grammar-constrained sampling may actually sample from an incorrect conditional distribution, but many sampling algorithms have been proposed to solve this problem~\cite{cars, mcmceasy, awrs}.}

Putting everything together, a constrained-sampling seed generator repeatedly produces syntactically valid samples via grammar-constrained decoding and collects them until $N$ samples are obtained that also satisfy the dialect’s semantic constraints.
\begin{definition}[Constrained-Sampling Seed Generator]
\label{def:constrained-sampling-gen}
Given a language model $M$, a constrained-sampling seed generator $\mathcal{G}$ for a dialect $\mathcal{D}$ produces $N$ samples $\{s_1,\ldots,s_N\}$ as follows:
\begin{enumerate}
  \item Obtain the dialect grammar $G_{\mathcal{D}}$ (\Cref{sec:grammar-extraction}).
  \item For $i=1\ldots N$: repeat $s_i \sim Q_{LM\mid G_{\mathcal{D}}}$ until $\varphi(s_i)$.
\end{enumerate}  
\end{definition}
   


Although $G_{\mathcal{D}}$ encodes only syntactic constraints, many MLIR dialect grammars also embed shallow semantic information (operand counts, attribute shapes, and type forms).  
Thus, a large fraction of validity checks are enforced directly during generation.
As we show in \Cref{sec:evaluation}, because grammatically invalid seeds are never produced, many samples are accepted by the compiler front end---i.e., they satisfy the semantic constraint $\varphi_\mathcal{D}$---drastically reducing the number of times the loop in item \textit{(2)} of \Cref{def:constrained-sampling-gen} is executed.  
Within the valid space \(L(G_{\mathcal{D}})\), the LM's priors continue to guide token choices, maintaining representativeness.  

For example, grammar-constrained sampling prevents generating the incomplete program in Figure~\ref{fig:gcd-comparison}b by recognizing that the \texttt{emitc.for} production requires three operands plus a region; tokens that would prematurely close the operation (e.g., \texttt{":"}) are masked out until all required components are present, forcing the LM to complete the structure.

Overall, grammar-constrained sampling transforms the LM from a high-variance, low-yield generator into an efficient seed generator.
The resulting generator satisfies all properties of a good seed generator (\Cref{def:seed-generator}):  
\emph{Dialect-agnostic}---it requires only grammars automatically derived from dialect definitions;  
\emph{Representative}---LM priors bias sampling toward realistic programs within the grammar's space; and  
\emph{Efficient}---yields hundreds of seeds per minute, enabling immediate large-scale coverage-guided fuzzing.

\section{Implementation}
\label{sec:implementation}



\name is implemented as a coverage-guided fuzzer built on top of \textsc{SynthFuzz}~\cite{SynthFuzz}, but with a new dialect-agnostic front end that automatically produces high-quality initial seeds.  
Two technical components underpin this design:
\begin{enumerate}[noitemsep]
    \item \textbf{Automatic Dialect Grammar Extraction:}  
    A pipeline that translates TableGen specifications into context-free grammars with principled fallbacks for custom formats (\Cref{sec:grammar-extraction}).

    \item \textbf{Grammar-Constrained LM Sampling:}  
    A seed generator that uses these grammars to constrain the outputs produced by a pre-trained LM, producing syntactically valid, diverse, and representative programs without any dialect-specific tuning (\Cref{sec:gcd-implementation}).
\end{enumerate}

The generated seeds are fed into SynthFuzz (\Cref{sec:fuzzing-loop}), which uses its mutation engine to adapt based on coverage signals, efficiently steering exploration toward high-value code regions.

\subsection{Automatic Grammar Extraction}
\label{sec:grammar-extraction}




\name automatically extracts (when possible) a context-free grammar (CFG) encoding the dialect's \emph{assembly syntax}, requiring no dialect-specific engineering.
The translation operates in two steps: 
(\textit{i}) mechanically translate TableGen specifications to grammar rules (\Cref{sec:tablegen-to-grammar});
(\textit{ii}) use LMs to infer grammar rules for custom formats specified in C++ outside TableGen, with a fallback to MLIR's generic format when inference fails (\Cref{sec:llm-to-rules}).

\subsubsection{Extracting from TableGen Specifications}
\label{sec:tablegen-to-grammar}


\name takes advantage of the fact that MLIR dialects are described in TableGen, a declarative language specifying operation syntax, operand/result constraints, and attribute schemas.  
For each TableGen operation definition, we extract the following information:
\begin{itemize}[noitemsep]
\item 
Operations with explicit \texttt{assemblyFormat} declarations directly encode their syntax (e.g., \texttt{"\$value attr-dict `:` type(\$result)"} for \texttt{emitc.literal} in Figure~\ref{fig:grammar-extraction}a).

\item 
Type constraints define valid types for operands and results (e.g., \texttt{Integer\-IndexOrOpaqueType} allows \texttt{index}, \texttt{i1}, \texttt{i8}, ..., \texttt{i128} types, as shown in Figure~\ref{fig:grammar-extraction}).

\item 
Attribute constraints specify metadata types (e.g.,  \texttt{I64ArrayAttr} for integer arrays).

\item 
Operations inherit format specifications from parent classes through TableGen's class system.
\end{itemize}

These specifications are translated into context-free grammar rules using the Lark grammar format, which is supported by many grammar-constrained sampling LM libraries.
Figure~\ref{fig:grammar-extraction}.b shows example extracted rules for the \texttt{EmitC} dialect. The resulting grammar $\mathcal{G}_{\mathcal{D}}$ extends a core MLIR template grammar (\Cref{app:grammar-template}) with dialect-specific productions.


\subsubsection{Handling Custom Assembly Formats.}
\label{sec:llm-to-rules}

In MLIR, operations with \texttt{hasCustomAssemblyFormat = 1} specify syntax in C++ rather than TableGen (see \texttt{emitc.for} in Figure~\ref{fig:grammar-extraction}a), so their format cannot be extracted automatically.  
To extract grammar rules for this type of operation, \name uses a two-tier fallback strategy.
First, it searches test files for occurrences of the operation and infers a grammar rule using an LM. 
The LM is prompted with multiple usage examples from tests, infers their common syntactic structure, and produces a candidate grammar rule that is then validated against the examples.
For the \texttt{emitc.for} operation from Figure~\ref{fig:grammar-extraction}a, examples are found in test files showing usage like
\texttt{emitc.for \%i = \%lb to \%ub step \%s : index \{ emitc.yield \}}, and
the LM infers the rule shown in Figure~\ref{fig:grammar-extraction}     (highlighted in \textcolor{magenta}{magenta}), capturing the \texttt{SSA\_ID = ssa\_use to ssa\_use step ssa\_use} pattern.

Second, if no usage examples exist for the operation, or inference produces a rule that fails validation on the usage example, we mark the operation for generation in the MLIR \emph{generic format}, which every operation supports by construction.
Briefly, MLIR supports two equivalent representations: \textit{assembly format} (human-readable, dialect-specific syntax) and \textit{generic format} (uniform, machine-oriented syntax). 
This last format is uniform across all dialects and always syntactically valid.
While both formats support grammar-constrained generation, they differ in expressiveness:
Assembly format grammars capture dialect-specific syntax---type vocabularies (\texttt{IntegerIndexOrOpaqueType}), operation keywords (\texttt{to}, \texttt{step}), attribute schemas---enabling the LM to generate dialect-conforming programs during constrained generation.
Generic format grammars only enforce MLIR's universal structure (\texttt{"dialect.op"(...) : (...) -> (...)}) without encoding which operations, types, or attributes belong to the dialect.
Consequently, generic format generation requires post-hoc filtering: the LM may generate syntactically valid but dialect-inappropriate operations (e.g., \texttt{arith.addi} in an \texttt{EmitC}-only program), leading to invalid programs.

\begin{wrapfigure}{r}{0.4\linewidth}
\vspace{-1em}
\begin{lstlisting}[language=MLIR, basicstyle=\footnotesize\ttfamily]
"emitc.for"(%lb, %ub, %s) ({
  ^bb0:
    "emitc.yield"() : () -> ()
}) : (index, index, index) -> ()
\end{lstlisting}
\caption{Generic format.}
\label{fig:generic}
\vspace{-1em}
\end{wrapfigure}
For our running example, if the LM had generated an invalid rule that does not match the examples found in the test files for \texttt{emitc.for}, we would mark \texttt{emitc.for} for generic format generation only as shown in \Cref{fig:generic}.
Since MLIR prohibits mixing formats in a single file, if any operation lacks an extractable assembly grammar, the entire program must use generic format.

\subsection{Grammar-Constrained Seed Generation}
\label{sec:gcd-implementation}

Given the extracted grammar $\mathcal{G}_{\mathcal{D}}$, \name generates seed programs using a pre-trained LM equipped with grammar-constrained sampling.
We note that grammar-constrained sampling libraries support any open-source model that exposes its token logits---i.e., the probabilities associated with individual tokens. Concretely, \name uses vLLM~\cite{vllm} with the Guidance~\cite{GuidanceAI2023} backend.

\subsubsection{Assembly vs.\ Generic Format.}
When grammar extraction does not fall back to the generic format, the extracted grammars describe assembly format since TableGen specifies assembly syntax. 
However, the generic format exposes bugs that assembly parsers mask through automatic corrections.
For example, assembly parsers may infer missing type annotations or silently fix malformed attributes, preventing such inputs from reaching the verifier where bugs lurk.

We thus adopt a two-stage strategy:
First, if a grammar over assembly has been extracted, we use grammar-constrained sampling with $\mathcal{G}_{\mathcal{D}}$ to generate seeds. 
Second, we convert seeds to generic format using MLIR's standard \texttt{mlir-opt} tool with the \texttt{--mlir-print-op-generic} flag. Generic format directly reflects in-memory IR structure, bypassing custom parser corrections that mask verifier bugs.
If a grammar over assembly was not extracted, we simply grammar-constrained sample over the generic grammar (with all the dialect operations).


This could also explain why assembly format achieves higher validity rates (Table~\ref{tab:validity}): dialect-specific constraints reduce the LM's search space to valid combinations. Still, the generic format's universal structure provides a guaranteed fallback when assembly grammars cannot be extracted.

\subsubsection{Prompt Construction.}
When test files exist, we automatically build few-shot prompts with 4 example programs selected to span diverse complexities (measured by AST depth and operation count). This encourages structural diversity in generated seeds.

When \emph{no examples exist}---in low-resource dialects---we construct zero-shot prompts from:
\begin{itemize}[noitemsep]
    \item A generic MLIR module skeleton (e.g., ``You are generating an MLIR program for the \texttt{<dialect>} dialect. Generate a valid module with multiple operations.''),
    \item 1–2 syntactic scaffolds---template code snippets showing basic structure like region syntax (\texttt{\{\textasciicircum bb0: ... \}}), block organization, and type declarations,
    \item A description of available operations and their signatures extracted from TableGen (e.g., ``\texttt{emitc.literal}: takes a string attribute, returns an EmitCType'').
\end{itemize}

\Cref{app:prompts} shows complete examples of both few-shot and zero-shot prompts.
We stress that all prompts are generated automatically.

\subsubsection{Semantic Filtering}
\label{sec:semantic-filtering}

When grammar-constrained generation produces syntactically valid candidates, \name applies semantic filtering to avoid producing invalid outputs (step 2 of Definition~\ref{def:constrained-sampling-gen}) .
While constrained sampling increases semantic validity, not all syntactically valid programs pass semantic constraints---deeper constraints like SSA form, type consistency, and dominance cannot be enforced by context-free grammars.
In our implementation, we decide to retains both \emph{strictly valid} programs (passing $\varphi_\mathcal{D}$) and what we call \emph{soft valid} programs, which are not strictly valid but judged to be semantically plausible by the LM.
Soft valid programs provide structural patterns that mutation operators can repair during fuzzing---i.e., soft valid samples are very effective in practice.
We validate this design empirically in \Cref{subsec:seed_validity}; the LM judging mechanism is detailed in \Cref{app:llm-judge}.
To summarize, programs are classified into three categories: \textbf{Strictly valid}: Pass MLIR's semantic verifier $\varphi_{\mathcal{D}}$ (SSA form, type consistency, dominance, etc.); 
\textbf{Soft valid}: Fail $\varphi_{\mathcal{D}}$ but judged semantically plausible by the LM (threshold $\tau=7$; see \Cref{app:llm-judge}); and
\textbf{Rejected}: Fundamentally malformed programs.



\subsection{Coverage-Guided Fuzzing Loop}
\label{sec:fuzzing-loop}

The seed corpus $S$ generated in \Cref{sec:gcd-implementation} is fed into our coverage-guided fuzzing engine, which integrates SynthFuzz~\cite{SynthFuzz} as its core mutation engine.
SynthFuzz is a state-of-the-art, grammar-aware fuzzer that automatically infers and synthesizes context-dependent mutations from a seed corpus. Unlike traditional coverage-guided fuzzers that use random bit-flips, SynthFuzz analyzes syntactic patterns in the seeds to learn complex, valid mutations (e.g., statement insertion, operand substitution, type changes). This makes it well-suited for structured languages like MLIR.


After feeding the initial seed corpus $S$, \name wraps SynthFuzz in a coverage-guided feedback loop, which at each iteration:
\rone Selects a seed from the corpus, prioritizing inputs that recently increased code coverage.
\rtwo Invokes SynthFuzz to apply a synthesized, context-aware mutation to the seed.
\rthree Executes the mutated program on the project under test using \texttt{mlir-opt}.
\rfour Records any crashes or verifier failures as potential bugs.
\rfive If the mutant increases coverage, it is added to the corpus for future iterations.
In summary, our grammar-based seeds ensure broad initial coverage and provide the foundational patterns for SynthFuzz to learn from, while the coverage-guided loop steers these intelligent mutations towards deeper program exploration. 

\section{Evaluation}
\label{sec:evaluation}

We evaluate \name across six MLIR-based compiler projects, discovering 88 previously unknown bugs and demonstrating substantial improvements in both bug discovery and code coverage compared to existing approaches. Our evaluation answers:

\begin{itemize}
    \item \textbf{RQ1 (Bug Discovery):} Can \name discover real bugs in production compilers, and how does discovery timing compare to baselines?
    \item \textbf{RQ2 (Seed Quality):} How does \name's seed generation improve fuzzer effectiveness compared to existing seed sources?
\end{itemize}

\subsection{Experimental Setup}

\subsubsection{Target Projects.} 
We evaluate on six MLIR-based projects representing diverse domains and maturity levels (full statistics in \Cref{app:targets}):
\textbf{MLIR} (base infrastructure, 45 dialects, 964 ops),
\textbf{Torch-MLIR} (PyTorch import, 7 dialects, 215 ops),
\textbf{IREE} (ML runtime, 38 dialects, 659 ops),
\textbf{CIRCT} (hardware, 20 dialects, 114 ops),
\textbf{Triton} (GPU kernels, 13 dialects, 76 ops), and
\textbf{HEIR} (cryptography, 14 dialects, 75 ops).
These span mature ecosystems with extensive tests (MLIR: 2,000; IREE: 1,100), high-profile projects with limited testing (Triton: 17,500 stars, 80 tests), and emerging dialects where testing lags development (HEIR: 600 stars, 170 tests).
MLIR serves as the baseline, hosting core dialects upon which other projects build.

\subsubsection{Baseline Fuzzers.}
The only \textit{dialect-agnostic} tools we can compare against across all six projects are:
\textbf{Grammarinator}~\cite{grammarinator}---a grammar-based fuzzer that generates and evolves existing test cases using context-free grammars through sampling, mutation, and recombination---and \textbf{SynthFuzz}~\cite{SynthFuzz}---a Grammar-based fuzzer that automatically learns and applies context-dependent, parameterized mutations from existing test cases.

For the base MLIR project, we can additionally compare against MLIR-specific fuzzers (extending these approaches to other projects would require substantial manual effort per dialect, which is what this paper tries to avoid):  \textbf{MLIRSmith}~\cite{MLIRSmith}---a hand-engineered type-consistent generator with manual rules per dialect---and
\textbf{MLIRod}~\cite{MLIRod}---a coverage-guided fuzzer that uses operation dependency graphs and directed mutations to guide test generation.

We excluded \textbf{FLEX}~\cite{FLEX} from our evaluation as it does not satisfy our design goal of enabling rapid, dialect-agnostic fuzzing. FLEX requires iterative fine-tuning before fuzzing can begin: the paper reports 30 training iterations totaling 45-60 GPU-hours per dialect on RTX 4090 GPUs. Applying this to our six-project evaluation would require over 270 GPU-hours of upfront training. Additionally, FLEX assumes the availability of substantial training data (15,344 seed programs in the paper), which is unavailable for low-resource dialects.

\subsubsection{Evaluation Design.}
Our evaluation uses two complementary settings:

\textbf{End-to-End Fuzzer Comparison.} We compare \name as a complete system against baseline fuzzers using default configurations. Each fuzzer uses its native seed strategy: \name uses 1000 generated seeds via grammar-constrained LM sampling over the project's existing test suite, while baselines use only the project's existing test suite.

\textbf{Seed Quality Isolation.} To isolate seed generation from the effect of different fuzzing algorithms, we test baseline fuzzers with three configurations:  \textbf{Original}: Seeds from the project's existing test suite;  \textbf{Original + Random-CFG (1K)}: Original corpus plus 1,000 random  samples generated using Grammarinator;  and \textbf{Original + \name (1K)}: Original corpus plus 1,000 \name seeds.

\subsubsection{Evaluation Metrics and Procedure.}
We assess fuzzer effectiveness along two dimensions: \textbf{Line Coverage}: Number of compiler implementation lines exercised during fuzzing, and \textbf{Bug Discovery}: Cumulative unique crashes over time, deduplicated by stack trace. All bugs are manually triaged to eliminate false positives and reported to maintainers.

For each configuration, we conduct five independent 24-hour fuzzing trials to account for randomness, reporting average coverage and total unique bugs across trials. In total, the study comprises 100 end-to-end experiments (five projects × three tools × five trials, plus one project × five tools × five trials) and 195 seed-quality-isolation experiments (six projects × three seed configurations × three tools × five trials, excluding 75 that overlap with the end-to-end setting), amounting to 7,080 fuzzer-hours. 
All evaluations are executed on dedicated Ubuntu 22.04 LTS nodes equipped with Intel Xeon Gold 6230 processors (2.10 GHz, 10 cores, 20 threads allocated), 384 GB of memory, and—where applicable---a single NVIDIA RTX A6000 accelerator. The software environment consists of Python 3.10.12, PyTorch 2.8.0 with CUDA 12.8, Transformers 4.55.4, and llguidance 0.7.30, ensuring consistent and reproducible execution across all workloads.

\name employs Qwen-2.5-7B-Coder-Instruct~\cite{qwen}---a relatively small, open-source model that can run locally without API costs or cloud infrastructure. We use temperature 1.0, top-p 1.0, and 4 few-shot examples per dialect. We deliberately chose a mid-sized model (7B parameters) rather than larger alternatives to demonstrate that \name's effectiveness stems from its grammar-constrained approach rather than model scale. 
Complete implementation details and hyperparameters appear in~\Cref{app:hyperparams,app:models,app:overview}.


\subsection{RQ1: Can \name discover real bugs in production compilers, and how does discovery timing compare to baselines?}
\subsubsection{Bug Discovery Effectiveness.}
\begin{figure}[t]
\centering
\includegraphics[width=\linewidth]{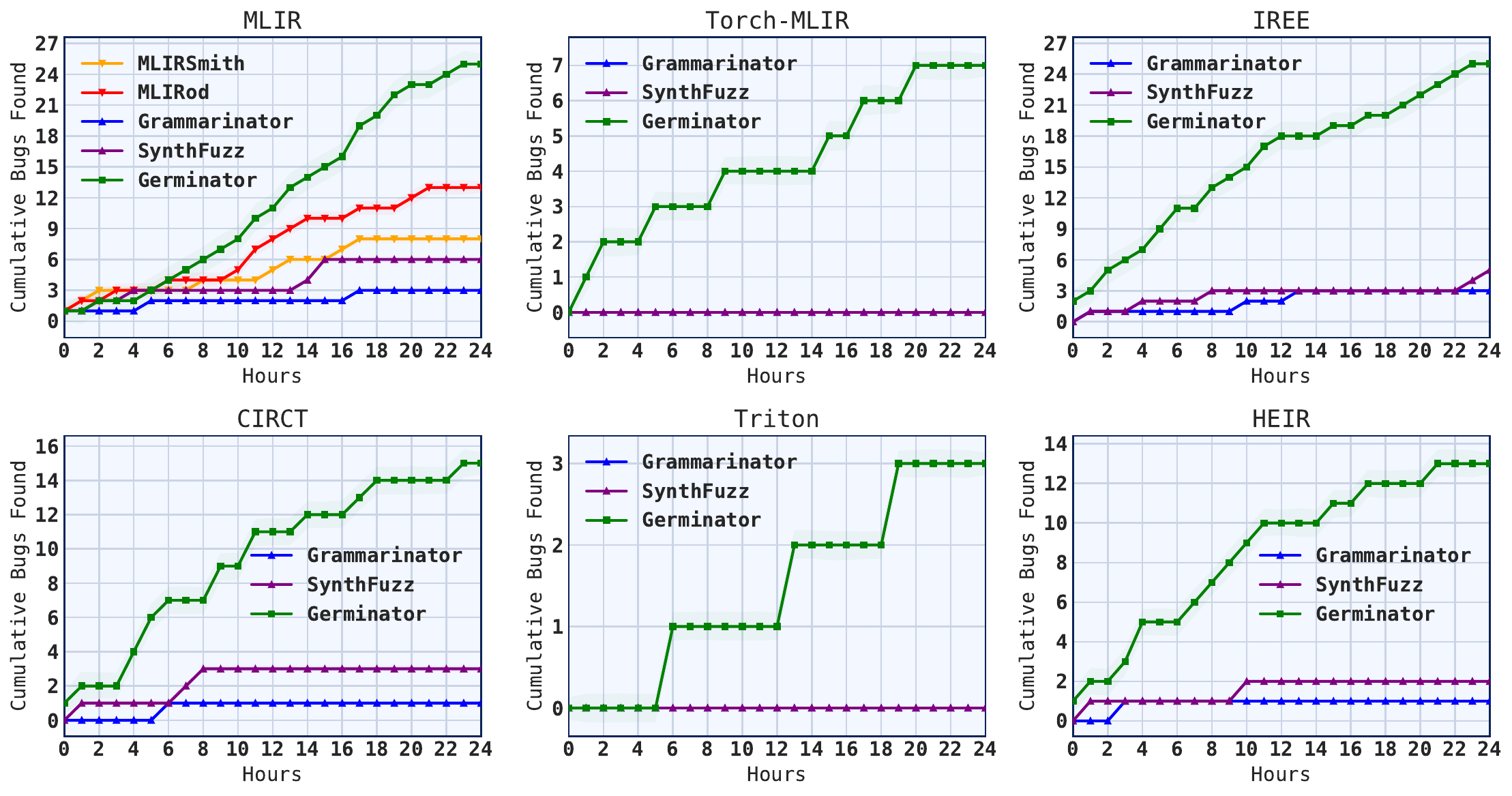}
\caption{Cumulative bug discovery over 24 hours across all six target projects. \name discovers substantially more bugs and finds them earlier than baselines. In Torch-MLIR and Triton, baselines discover zero bugs while \name finds 7 and 3 bugs respectively. Shaded regions indicate 95\% bootstrapped confidence intervals across 5 trials.}
\label{fig:bugs-over-time}
\end{figure}

Figure~\ref{fig:bugs-over-time} shows cumulative unique bugs discovered over 24-hour campaigns across all projects. Each crash is deduplicated by stack trace and manually triaged to confirm genuine compiler defects.

\name substantially outperforms baselines in bug discovery. 
~\Cref{app:bug-counts} summarizes final counts after 24 hours. Across all projects, \name discovers 88 unique bugs compared to 8 for Grammarinator and 16 for SynthFuzz---a 5.5$\times$ cumulative improvement over the best baseline. Critically, in  Torch-MLIR and Triton, baseline fuzzers found zero bugs while \name discovered 7 and 3, respectively.




\subsubsection{Bug Characteristics.}
We manually analyzed all 88 bugs discovered by 
\begin{wrapfigure}{r}{0.43\textwidth}
\vspace{-1.6em}
\caption{Bug types discovered by \name.}
\centering
\begin{tabular}{lrr}
\toprule
\textbf{Bug Category} & \textbf{\#} & \textbf{\%} \\
\midrule
Type system violations & 26 & 29.5\% \\
Dialect conversion errors & 20 & 22.7\% \\
Optimization pass crashes & 16 & 18.2\% \\
Verifier inconsistencies & 12 & 13.6\% \\
Lowering failures & 8 & 9.1\% \\
Memory safety issues & 6 & 6.8\% \\
\bottomrule
\end{tabular}
\label{tab:bug-types}
\vspace{-0.6em}
\end{wrapfigure}
\name to understand their nature and severity.
\Cref{tab:bug-types} categorizes bugs
 by root cause and affected compiler component.

\begin{itemize}
\item \textbf{Type system violations (29.5\%)} dominated the bug reports, revealing fundamental challenges in MLIR's complex type hierarchy and casting operations across diverse dialects.
\item \textbf{Dialect conversion errors (22.7\%)} highlighted interoperability issues, particularly during legalization patterns and multi-dialect lowering pipelines.
\item The distribution shows \name effectively stresses core MLIR components, with over 70\% of bugs affecting critical compiler infrastructure rather than peripheral features.
\end{itemize}

\subsubsection{Bug Confirmation and Reporting.}
We reported all 88 bugs to project maintainers. As of submission, 40 have been confirmed and 14 have been fixed, demonstrating that \name discovers genuine, previously unknown defects rather than test harness artifacts. The remaining bugs await triage. 

To validate the practical relevance of discovered bugs, we shared our findings with MLIR maintainers and dialect developers on the official LLVM Discord server.
We received permission to report their feedback.
A core MLIR contributor (Developer A) reviewed our bug reports and assessed them as substantive: \textit{"These all seem substantive to me"} (August 2025).
Two additional maintainers emphasized that our findings aligned with established quality expectations for MLIR infrastructure.
Regarding verifier crashes, Developer B noted: \textit{"Verifiers should handle invalid IR without crashing. I think the bugs you've found there are great!"} (August 2025).
These responses confirm that \name's bug discoveries address real robustness issues and that the tool generates test cases exercising important corner cases that existing test suites miss.


{\setlength{\fboxsep}{8pt}%
\noindent\fbox{\parbox{0.95\linewidth}{%
\textbf{RQ1 Summary:} \name discovers 5.5$\times$ more bugs than baselines (88 vs 16), finds bugs in projects where baselines discover zero, and receives confirmation from MLIR maintainers that findings are substantive and address real robustness issues.
}}}

\subsection{RQ2: How does \name's seed generation improve fuzzer effectiveness?}

\subsubsection{Seed Validity.}
\label{subsec:seed_validity}

We empirically validate the three-level validity classification from Definition~\ref{def:constrained-sampling-gen} and our semantic filtering implementation (\Cref{sec:semantic-filtering}).
Table~\ref{tab:validity} reports validity rates and time to produce 1,000 soft valid seeds---the seed budget used in all evaluations.

Grammar-constrained generation ensures 100\% syntactic validity by construction, enabling high soft validity.
Unconstrained LM generation achieves only 4-5\% syntactic validity, constraining soft validity to 2.8-3.5\%.
In contrast, \name achieves 36.2\% (assembly) and 28.8\% (generic) soft validity---a $\sim$14$\times$ improvement---by ensuring all outputs can be judged for semantic plausibility.

\name's strict validity (14.1\% assembly, 6.6\% generic) represents 100$\times$ improvement over random sampling and 10$\times$ over unconstrained generation.
Random-CFG ensures syntactic validity but achieves only 0.15\% soft validity due to unrealistic combinations from uniform sampling.
\name generates 1,000 soft valid seeds in $\sim$6 minutes, 16-20$\times$ faster than unconstrained generation (86-107 minutes).


\begin{table}[t]
\caption{Seed validity across generation strategies, averaged over all six projects. }
\small
\centering
\begin{tabular}{lrrrrr}
\toprule
\textbf{Strategy} & \textbf{Format} & \textbf{Syntax Valid} & \textbf{Strictly Valid} & \textbf{Soft Valid} & \textbf{Time (1K seeds)} \\
\midrule
\name & Assembly & 100\% & 14.1\% & 36.2\% & 6 min \\
\name & Generic & 100\% & 6.6\% & 28.8\% & 6 min \\
\midrule
LM (rejection) & Assembly & 4.2\% & 1.4\% & 2.8\% & 107 min \\
LM (rejection) & Generic & 5.1\% & 2.1\% & 3.5\% & 86 min \\
\midrule
Random-CFG & Assembly & 100\% & 0.01\% & 0.015\% & <1 min \\
Random-CFG & Generic & 100\% & 0.01\% & 0.012\% & <1 min \\
\bottomrule
\end{tabular}
\label{tab:validity}
\end{table}


\subsubsection{Grammar Extraction Success.}


Across all six projects, \name successfully extracts grammars for 88.9\% of the operations spanning 137 dialects (More details in Appendix~\ref{app:grammar-extraction}).
For 82.8\% of the operations following standard TableGen patterns (declarative assembly format) require no manual intervention (e.g., \texttt{arith.addi}, \texttt{emitc.literal}, \texttt{func.call}).
For 6.1\% of the operations with non-standard TableGen definitions, \name could infer rules using an LM base on existing examples and available information (e.g., \texttt{emitc.for}, \texttt{gpu.func}).

The 11.1\% of the operations that cannot be automatically extracted are due to: (a) missing/malformed TableGen specs, (b) pure C++ definitions without declarative specs, or (c) highly irregular syntax requiring dialect-specific parsing. These represent edge cases in mature dialects---often deprecated operations or internal details not intended for direct construction.
Even for the 11.1\% where extraction fails, \name's LM component can generate syntactically valid programs by falling back to a generic MLIR grammar.

\subsubsection{Line Coverage.}

\begin{figure}[t]
\centering
\includegraphics[width=\linewidth]{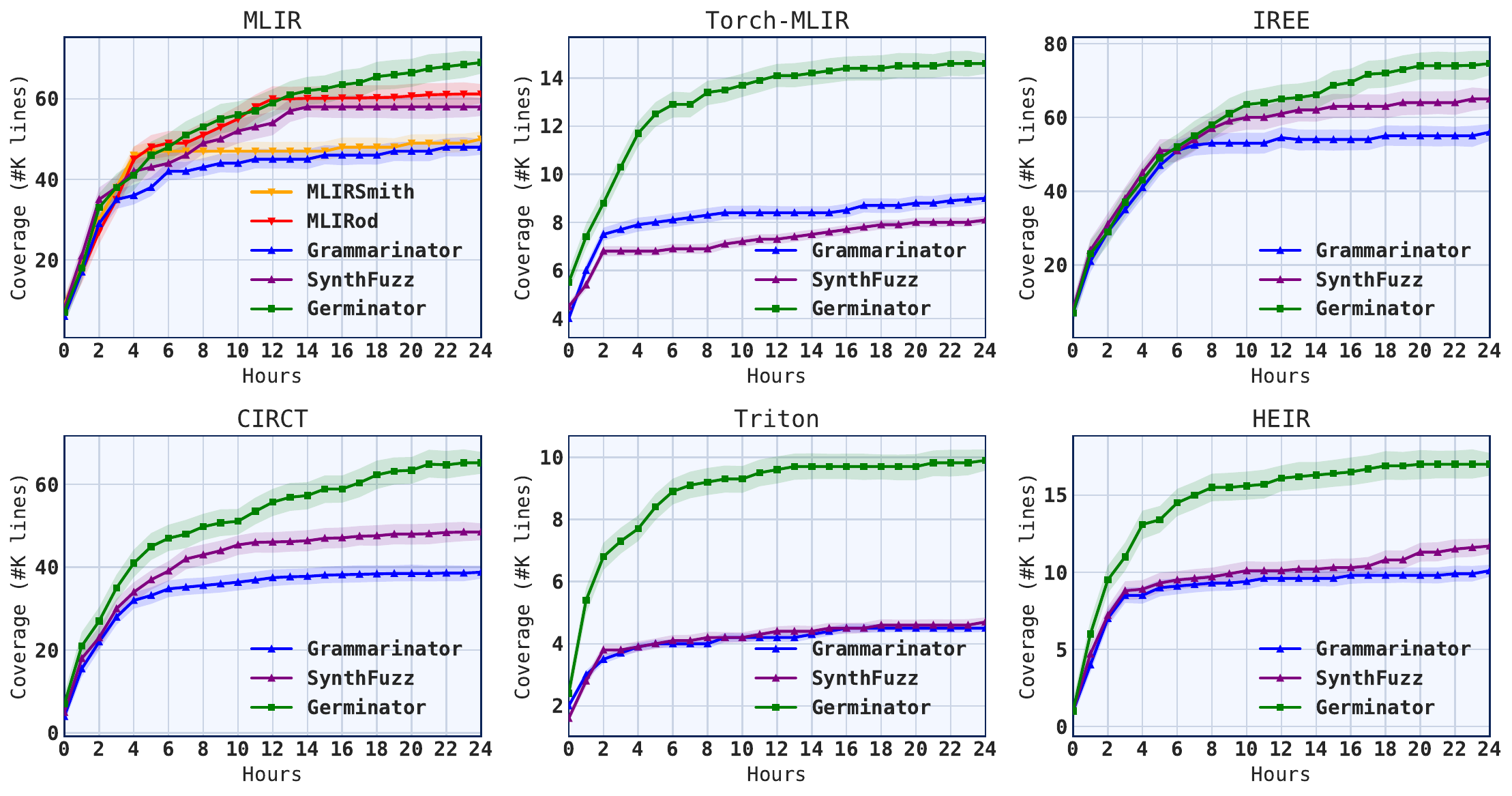}
\caption{Line coverage evolution over 24 hours across all six projects. Performance gaps are most pronounced in projects with limited test infrastructure. Shaded regions indicate 95\% bootstrapped confidence intervals across 5 trials.}
\label{fig:coverage-over-time}
\end{figure}

Figure~\ref{fig:coverage-over-time} presents 24-hour line coverage trajectories across our six projects in the end-to-end comparison. \name consistently achieves higher final coverage and faster growth compared to baselines, with noticeable gains in low-resource dialects.

Across all projects, \name achieves 10-120\% higher final coverage than the best baseline. Advantages are most pronounced in limited-test-infrastructure projects: Triton (80 tests, 2.2$\times$ improvement), HEIR (170 tests, 1.45$\times$), CIRCT (650 tests, 1.34$\times$). Even in mature IREE (1,100 tests), \name achieves 33\% higher coverage (74.6 KLOC vs 65.0 KLOC).

Divergent trajectories reveal fundamental differences in exploration strategies. Baseline approaches---Grammarinator's grammar-based enumeration and SynthFuzz's learned mutations---operate without coverage feedback. While they maintain diversity through mutation, they lack mechanisms to guide generation toward unexplored regions. 

\name incorporates coverage-guided feedback into its fuzzing loop. This explains both higher final coverage and delayed saturation---the system actively learns which regions remain unexplored and steers generation accordingly rather than relying on static diversity heuristics.

Beyond absolute gains, \name demonstrates superior exploration efficiency through faster early growth. Within 4 hours, \name typically reaches coverage levels requiring 15-20 hours for baselines or never attained. In Torch-MLIR, \name reaches 11.7 KLOC in 4 hours---a threshold baselines approach only after 20+ hours. This early advantage stems from \name's ability to generate diverse, structurally complex seeds exposing deep compiler paths immediately rather than gradually building complexity through mutation of simple initial seeds.

\subsubsection{Seed Quality Isolation.}
To isolate \name's seed generation contribution from its fuzzing algorithm, we evaluate how different seed sources affect baseline fuzzer performance. Table~\ref{tab:seed-quality} compares 24-hour line coverage achieved by Grammarinator and SynthFuzz when augmented with different seed corpora.

\begin{table}[t]
\caption{Impact of seed corpus augmentation on baseline fuzzer coverage. All configurations include original test suite plus 1,000 additional seeds from specified strategy. \name seeds consistently improve coverage by 20-55\% over original seeds and 15-55\% over random grammar sampling.}
\small
\centering
\begin{tabular}{llrrr}
\toprule
\textbf{Project} & \textbf{Fuzzer} & \textbf{Original} & \textbf{+Random-} & \textbf{+\name} \\
 & & \textbf{(KLOC)} & \textbf{CFG (KLOC)} & \textbf{(KLOC)} \\
\midrule
\multirow{2}{*}{MLIR} & Grammarinator & 48.1 & 48.2 & 49.1 \\
 & SynthFuzz & 58.1 & 58.1 & 59.4 \\
\midrule
\multirow{2}{*}{Torch-MLIR} & Grammarinator & 9.0 & 6.7 & 9.1 \\
 & SynthFuzz & 8.1 & 6.5 & 9.8 \\
\midrule
\multirow{2}{*}{IREE} & Grammarinator & 56.0 & 56.2 & 59.5 \\
 & SynthFuzz & 65.0 & 65.2 & 67.2 \\
\midrule
\multirow{2}{*}{CIRCT} & Grammarinator & 38.8 & 38.1 & 44.7 \\
 & SynthFuzz & 48.5 & 49.0 & 53.9 \\
\midrule
\multirow{2}{*}{Triton} & Grammarinator & 4.5 & 3.3 & 6.8 \\
 & SynthFuzz & 4.7 & 3.9 & 7.2 \\
\midrule
\multirow{2}{*}{HEIR} & Grammarinator & 10.1 & 10.1 & 12.5 \\
 & SynthFuzz & 11.7 & 11.5 & 13.3 \\
\bottomrule
\end{tabular}
\label{tab:seed-quality}
\end{table}

Augmenting baseline fuzzers with \name seeds yields substantial coverage improvements. Compared to using only original tests, adding 1,000 \name seeds improves Grammarinator by 3-41\% and SynthFuzz by 4-43\%. 

Interestingly, Random-CFG frequently \textit{degrades} performance---reducing coverage by 1-27\% in Torch-MLIR, CIRCT, and Triton. 
The seed quality gap is most pronounced in low-resource settings. In Triton, \name seeds
enable Grammarinator to reach 6.8 KLOC whereas Random-CFG seeds degrade the coverage to 3.3 KLOC---a 106\% gap between the two.
This result reveals that na\"ive seed augmentation can harm effectiveness: low-quality seeds dilute the corpus and waste mutation budget on unpromising starting points. 


Even in mature projects with extensive tests, \name seeds provide meaningful gains. In IREE (1,100 tests), \name improves Grammarinator coverage by 6\% (59.5 vs 56.0 KLOC) and SynthFuzz by 4\% (67.2 vs 65.0 KLOC), while Random-CFG provides negligible benefit (<1\%). This suggests handwritten test suites, despite their size, can exhibit structural biases limiting mutation-based exploration. \name can generate qualitatively different programs exposing under-tested paths even when thousands of existing tests are available.

The consistent advantage of \name seeds over both original tests and random generation demonstrates that effective seed quality requires more than syntactic validity---it demands semantic diversity and structural complexity that LM guidance can reliably produce at scale.

{\setlength{\fboxsep}{8pt}%
\noindent\fbox{\parbox{0.95\linewidth}{%
\textbf{RQ2 Summary:} \name achieves 14$\times$ higher soft validity than unconstrained generation, extracts grammars for 88.9\% of operations, reaches 10–120\% higher coverage than baselines, and improves baseline fuzzer coverage by 20–55\% when used as seed source alone.
}}}

\section{Related Work}
\label{sec:related}

We have discussed grammar-based fuzzing, learning-based fuzzing, hand-engineered MLIR fuzzers, and their dialect-(in)effectiveness in \Cref{sec:why-existing-fail}.  
Here we situate \name within broader research on fuzzing, program generation, and compiler testing.

\paragraph{Coverage-Guided Fuzzing.}
Coverage-guided fuzzing originated with AFL++~\cite{AFLpp} and has since evolved through improved mutation heuristics, energy allocation, greybox feedback, and ensemble techniques~\cite{AFLFast, FairFuzz, GreyOne, EnFuzz}.  
Grammar-guided fuzzers such as Grammarinator~\cite{grammarinator} and LangFuzz~\cite{LangFuzz} incorporate syntactic structure to avoid invalid inputs, while tools such as Skyfire~\cite{SkyFire} use probabilistic models to capture program distributions.  
Our work brings coverage-guided fuzzing to extensible compiler frameworks by making seed generation dialect-agnostic: \name automatically extracts grammars from dialect specifications, eliminating the need for manual grammar authoring for each new dialect.

\paragraph{Compiler and IR Fuzzing.}
Fuzzing compiler intermediate representations (IRs) has a rich history.  Early work such as Csmith~\cite{regehr2011} generated well-defined C programs to expose miscompilations.  EMI~\cite{Le12EMI} used equivalence modulo inputs to discover semantic inconsistencies.  Other approaches target LLVM IR specifically, such as the ``Alive'' family~\cite{alive, alive2}, superoptimizer-guided fuzzing~\cite{Souper}, or YARPGen~\cite{yarpgen} for generating random C/C++ programs to test compiler optimizations. For MLIR specifically, MLIRSmith~\cite{MLIRSmith} and MLIRod~\cite{MLIRod} implement hand-engineered fuzzers for specific dialects, while recent work has explored directed testing strategies~\cite{MLIRTracer}. 
Our work complements these approaches and enables dialect-agnostic and dialect-effective IR fuzzing across \emph{all} MLIR dialects without expert intervention.

\paragraph{Grammar Extraction.} 
Prior work on automatic grammar extraction falls into two categories: dynamic approaches that infer grammars from execution traces~\cite{ZellerLearnGrammar}, and static approaches that extract grammars from documentation~\cite{Polyglot}. Because MLIR dialects are defined via TableGen specifications that explicitly encode syntax through \texttt{assemblyFormat} strings---providing a declarative, machine-readable source of truth---our approach exploits this structure: rather than inferring approximate grammars from behavior, we extract precise grammars directly from TableGen definitions, ensuring consistency with the actual parser.
The primary technical challenge is handling the subset of operations with custom C++ formatters (\Cref{sec:grammar-extraction}), which we address using LMs.

\paragraph{Language Models for Program Generation.}
Recent work has explored using LMs to generate programs for testing and verification.  
FLEX~\cite{FLEX} uses LMs to generate test inputs for fuzzing language interpreters.
Other approaches leverage LMs to synthesize unit tests~\cite{LLMTest}, or guide symbolic execution~\cite{NeuralGuided}.
Learn\&Fuzz~\cite{LearnFuzz} pioneered combining neural models with mutation-based fuzzing, while recent work applies large language models directly to compiler testing~\cite{LLMDLFuzz}.

\name differs in two key respects:  
First, we enforce syntactic validity through grammar-constrained decoding using grammars extracted automatically from TableGen specifications, eliminating invalid programs by construction rather than through post-hoc filtering. Second, we use the LM only for seed generation within a coverage-guided fuzzing loop---the fuzzer then mutates these seeds to explore the program space---rather than relying on the LM for all test generation. 

\paragraph{Grammar-Constrained Sampling.}
Grammar-constrained sampling has been explored in NLP and code generation~\cite{geng-etal-2023-grammar, dong2023codepad, poesia2022synchrome, scholak2021picard, park2025flexible, GuidanceAI2023,parkGad}, including for producing structured JSON/XML or type-safe code.  While several sampling algorithms with different tradeoffs exist, our implementations uses the \texttt{guidance} grammar-constrained decoding library due to its efficiency~\cite{GuidanceAI2023}. 

\section{Conclusion}
\label{se:conclusion}

We showed that large language models can serve as effective oracles for \emph{bootstrapping} coverage-guided fuzzers for dialects in extensible languages: when combined with grammars automatically extracted from dialect specifications, language models produce small sets of representative, valid seeds that enable dialect-effective fuzzing without manual engineering.  
Using this approach, our tool \name uncovered 88 previously unknown bugs across MLIR dialects.

In terms of limitations, \name depends on information available in TableGen: dialects with heavily custom C++ assembly formats fall back to MLIR's generic syntax, reducing seed representativeness. Additionally, \name does not perform automatic test case minimization: when bugs are found, manual effort is required to distill reproducible examples for bug reports. Our approach enforces syntactic well-formedness but relies on the LM to satisfy deep semantic constraints such as SSA invariants (single static assignment, def-use consistency) or operation-specific preconditions encoded in dialect verifiers. Finally, our evaluation focuses on MLIR, whose declarative TableGen specifications make grammar extraction feasible; adapting \name to other frameworks would require similar declarative infrastructure.


\section*{Acknowledgments}
The project is supported in part by a Microsoft Faculty Fellowship; a UCSD JSOE Scholarship; NSF under grants CCF-2422214, CCF-2506134 and CCF-2446711; and the European Union, ERC grant (Project AT SCALE, 101179366).

We thank John Regehr for examining bug reports, and Maksim Levental for providing feedback on reported bugs.

Any opinions, findings, and conclusions or recommendations expressed in this publication are those of the authors, and do not necessarily reflect the views of the sponsoring entities.

Loris D'Antoni holds concurrent appointments as a Professor at the University of California San Diego and as an Amazon Scholar. This paper describes work performed at the University of California San Diego and is not associated with Amazon.

\bibliographystyle{ACM-Reference-Format}
\bibliography{main.bib}

@inproceedings{chengian2016,
author = {Sun, Chengnian and Le, Vu and Su, Zhendong},
title = {Finding compiler bugs via live code mutation},
year = {2016},
isbn = {9781450344449},
publisher = {Association for Computing Machinery},
address = {New York, NY, USA},
url = {https://doi.org/10.1145/2983990.2984038},
doi = {10.1145/2983990.2984038},
booktitle = {Proceedings of the 2016 ACM SIGPLAN International Conference on Object-Oriented Programming, Systems, Languages, and Applications},
pages = {849–863},
numpages = {15},
location = {Amsterdam, Netherlands},
series = {OOPSLA 2016}
}

@inproceedings{regehr2011,
author = {Yang, Xuejun and Chen, Yang and Eide, Eric and Regehr, John},
title = {Finding and understanding bugs in C compilers},
year = {2011},
isbn = {9781450306638},
publisher = {Association for Computing Machinery},
address = {New York, NY, USA},
url = {https://doi.org/10.1145/1993498.1993532},
doi = {10.1145/1993498.1993532},
booktitle = {Proceedings of the 32nd ACM SIGPLAN Conference on Programming Language Design and Implementation},
pages = {283–294},
numpages = {12},
location = {San Jose, California, USA},
series = {PLDI '11}
}

@inproceedings{xiao2019,
author = {Liu, Xiao and Li, Xiaoting and Prajapati, Rupesh and Wu, Dinghao},
title = {DeepFuzz: automatic generation of syntax valid C programs for fuzz testing},
year = {2019},
isbn = {978-1-57735-809-1},
publisher = {AAAI Press},
url = {https://doi.org/10.1609/aaai.v33i01.33011044},
doi = {10.1609/aaai.v33i01.33011044},
booktitle = {Proceedings of the Thirty-Third AAAI Conference on Artificial Intelligence and Thirty-First Innovative Applications of Artificial Intelligence Conference and Ninth AAAI Symposium on Educational Advances in Artificial Intelligence},
articleno = {129},
numpages = {8},
location = {Honolulu, Hawaii, USA},
series = {AAAI'19/IAAI'19/EAAI'19}
}

@article{chen2020,
author = {Chen, Junjie and Patra, Jibesh and Pradel, Michael and Xiong, Yingfei and Zhang, Hongyu and Hao, Dan and Zhang, Lu},
title = {A Survey of Compiler Testing},
year = {2020},
issue_date = {January 2021},
publisher = {Association for Computing Machinery},
address = {New York, NY, USA},
volume = {53},
number = {1},
issn = {0360-0300},
url = {https://doi.org/10.1145/3363562},
doi = {10.1145/3363562},
journal = {ACM Comput. Surv.},
month = feb,
articleno = {4},
numpages = {36}
}

@inproceedings{mlir,
  author={Lattner, Chris and Amini, Mehdi and Bondhugula, Uday and Cohen, Albert and Davis, Andy and Pienaar, Jacques and Riddle, River and Shpeisman, Tatiana and Vasilache, Nicolas and Zinenko, Oleksandr},
  booktitle={2021 {{IEEE/ACM}} International Symposium on Code Generation and Optimization (CGO)},
  title={{{MLIR}}: Scaling Compiler Infrastructure for Domain Specific Computation},
  year={2021},
  volume={},
  number={},
  pages={2-14},
  doi={10.1109/CGO51591.2021.9370308}
}

@misc{torch-mlir,
  title        = {Torch-MLIR},
  author       = {{LLVM Project}},
  howpublished = {\url{https://github.com/llvm/torch-mlir}},
  note         = {Accessed: 2025-09-04},
  year         = {2025},
}

@misc{circt,
  title        = {CIRCT: {C}ircuit {IR} Compilers and Tools},
  howpublished = {\url{https://circt.llvm.org/}},
  note         = {Accessed: 2025-09-04},
  year         = {2025}
}

@misc{iree,
  title        = {IREE: {I}ntermediate {R}epresentation {E}xecution {E}nvironment},
  howpublished = {\url{https://iree.dev/}},
  note         = {Accessed: 2025-09-04},
  year         = {2025}
}

@misc{triton,
  title        = {Triton: Open-source GPU Programming for Neural Networks},
  author       = {{OpenAI}},
  howpublished = {\url{https://openai.com/index/triton/}},
  year         = {2021},
  note         = {Accessed: 2025-09-04}
}

@misc{heir,
      title={HEIR: A Universal Compiler for Homomorphic Encryption}, 
      author={Asra Ali and Jaeho Choi and Bryant Gipson and Shruthi Gorantala and Jeremy Kun and Wouter Legiest and Lawrence Lim and Alexander Viand and Meron Zerihun Demissie and Hongren Zheng},
      year={2025},
      eprint={2508.11095},
      archivePrefix={arXiv},
      primaryClass={cs.CR},
      url={https://arxiv.org/abs/2508.11095}, 
}

@article{yarpgen,
author = {Livinskii, Vsevolod and Babokin, Dmitry and Regehr, John},
title = {Random testing for C and C++ compilers with YARPGen},
year = {2020},
issue_date = {November 2020},
publisher = {Association for Computing Machinery},
address = {New York, NY, USA},
volume = {4},
number = {OOPSLA},
url = {https://doi.org/10.1145/3428264},
doi = {10.1145/3428264},
journal = {Proc. ACM Program. Lang.},
month = nov,
articleno = {196},
numpages = {25}
}

@article{Marcozzi_2019,
   title={Compiler fuzzing: how much does it matter?},
   volume={3},
   ISSN={2475-1421},
   url={http://dx.doi.org/10.1145/3360581},
   DOI={10.1145/3360581},
   number={OOPSLA},
   journal={Proceedings of the ACM on Programming Languages},
   publisher={Association for Computing Machinery (ACM)},
   author={Marcozzi, Michaël and Tang, Qiyi and Donaldson, Alastair F. and Cadar, Cristian},
   year={2019},
   month=oct, pages={1–29} }

@inproceedings{grammarinator,
author = {Hodov\'{a}n, Ren\'{a}ta and Kiss, \'{A}kos and Gyim\'{o}thy, Tibor},
title = {Grammarinator: a grammar-based open source fuzzer},
year = {2018},
isbn = {9781450360531},
publisher = {Association for Computing Machinery},
address = {New York, NY, USA},
url = {https://doi.org/10.1145/3278186.3278193},
doi = {10.1145/3278186.3278193},
booktitle = {Proceedings of the 9th ACM SIGSOFT International Workshop on Automating TEST Case Design, Selection, and Evaluation},
pages = {45–48},
numpages = {4},
location = {Lake Buena Vista, FL, USA},
series = {A-TEST 2018}
}

@inproceedings{SynthFuzz,
author = {Limpanukorn, Ben and Wang, Jiyuan and Kang, Hong Jin and Zhou, Zitong and Kim, Miryung},
title = {Fuzzing MLIR Compilers with Custom Mutation Synthesis},
year = {2025},
isbn = {9798331505691},
publisher = {IEEE Press},
url = {https://doi.org/10.1109/ICSE55347.2025.00037},
doi = {10.1109/ICSE55347.2025.00037},
booktitle = {Proceedings of the IEEE/ACM 47th International Conference on Software Engineering},
pages = {217–229},
numpages = {13},
location = {Ottawa, Ontario, Canada},
series = {ICSE '25}
}

@inproceedings{MLIRod,
author = {Suo, Chenyao and Chen, Junjie and Liu, Shuang and Jiang, Jiajun and Zhao, Yingquan and Wang, Jianrong},
title = {Fuzzing MLIR Compiler Infrastructure via Operation Dependency Analysis},
year = {2024},
isbn = {9798400706127},
publisher = {Association for Computing Machinery},
address = {New York, NY, USA},
url = {https://doi.org/10.1145/3650212.3680360},
doi = {10.1145/3650212.3680360},
booktitle = {Proceedings of the 33rd ACM SIGSOFT International Symposium on Software Testing and Analysis},
pages = {1287–1299},
numpages = {13},
location = {Vienna, Austria},
series = {ISSTA 2024}
}

@inproceedings{MLIRSmith,
author = {Wang, Haoyu and Chen, Junjie and Xie, Chuyue and Liu, Shuang and Wang, Zan and Shen, Qingchao and Zhao, Yingquan},
title = {MLIRSmith: Random Program Generation for Fuzzing MLIR Compiler Infrastructure},
year = {2024},
isbn = {9798350329964},
publisher = {IEEE Press},
url = {https://doi.org/10.1109/ASE56229.2023.00120},
doi = {10.1109/ASE56229.2023.00120},
booktitle = {Proceedings of the 38th IEEE/ACM International Conference on Automated Software Engineering},
pages = {1555–1566},
numpages = {12},
location = {Echternach, Luxembourg},
series = {ASE '23}
}

@misc{FLEX,
      title={Interleaved Learning and Exploration: A Self-Adaptive Fuzz Testing Framework for MLIR}, 
      author={Zeyu Sun and Jingjing Liang and Weiyi Wang and Chenyao Suo and Junjie Chen and Fanjiang Xu},
      year={2025},
      eprint={2510.07815},
      archivePrefix={arXiv},
      primaryClass={cs.SE},
      url={https://arxiv.org/abs/2510.07815}, 
}

@misc{dong2023codepad,
      title={CodePAD: Sequence-based Code Generation with Pushdown Automaton}, 
      author={Yihong Dong and Xue Jiang and Yuchen Liu and Ge Li and Zhi Jin},
      year={2023},
      eprint={2211.00818},
      archivePrefix={arXiv},
      primaryClass={cs.SE},
      url={https://arxiv.org/abs/2211.00818}, 
}

@inproceedings{geng-etal-2023-grammar,
    title = "Grammar-Constrained Decoding for Structured {NLP} Tasks without Finetuning",
    author = "Geng, Saibo  and
      Josifoski, Martin  and
      Peyrard, Maxime  and
      West, Robert",
    editor = "Bouamor, Houda  and
      Pino, Juan  and
      Bali, Kalika",
    booktitle = "Proceedings of the 2023 Conference on Empirical Methods in Natural Language Processing",
    month = dec,
    year = "2023",
    address = "Singapore",
    publisher = "Association for Computational Linguistics",
    url = "https://aclanthology.org/2023.emnlp-main.674/",
    doi = "10.18653/v1/2023.emnlp-main.674",
    pages = "10932--10952"
}

@misc{scholak2021picard,
      title={PICARD: Parsing Incrementally for Constrained Auto-Regressive Decoding from Language Models}, 
      author={Torsten Scholak and Nathan Schucher and Dzmitry Bahdanau},
      year={2021},
      eprint={2109.05093},
      archivePrefix={arXiv},
      primaryClass={cs.CL},
      url={https://arxiv.org/abs/2109.05093}, 
}

@misc{poesia2022synchrome,
      title={Synchromesh: Reliable code generation from pre-trained language models}, 
      author={Gabriel Poesia and Oleksandr Polozov and Vu Le and Ashish Tiwari and Gustavo Soares and Christopher Meek and Sumit Gulwani},
      year={2022},
      eprint={2201.11227},
      archivePrefix={arXiv},
      primaryClass={cs.LG},
      url={https://arxiv.org/abs/2201.11227}, 
}

@misc{park2025flexible,
      title={Flexible and Efficient Grammar-Constrained Decoding}, 
      author={Kanghee Park and Timothy Zhou and Loris D'Antoni},
      year={2025},
      eprint={2502.05111},
      archivePrefix={arXiv},
      primaryClass={cs.CL},
      url={https://arxiv.org/abs/2502.05111}, 
}

@misc{emitc-docs,
  organization = {LLVM Project},
  title        = {EmitC Dialect — MLIR Documentation},
  howpublished = {\url{https://mlir.llvm.org/docs/Dialects/EmitC/}},
  year         = {2024},
  note         = {Accessed: 2025-10-09}
}

@misc{qwen,
      title={Qwen2.5-Coder Technical Report}, 
      author={Binyuan Hui and Jian Yang and Zeyu Cui and Jiaxi Yang and Dayiheng Liu and Lei Zhang and Tianyu Liu and Jiajun Zhang and Bowen Yu and Keming Lu and Kai Dang and Yang Fan and Yichang Zhang and An Yang and Rui Men and Fei Huang and Bo Zheng and Yibo Miao and Shanghaoran Quan and Yunlong Feng and Xingzhang Ren and Xuancheng Ren and Jingren Zhou and Junyang Lin},
      year={2024},
      eprint={2409.12186},
      archivePrefix={arXiv},
      primaryClass={cs.CL},
      url={https://arxiv.org/abs/2409.12186}, 
}

@inproceedings{dsp-mlir, series={LCTES ’25},
   title={DSP-MLIR: A Domain-Specific Language and MLIR Dialect for Digital Signal Processing},
   url={http://dx.doi.org/10.1145/3735452.3735527},
   DOI={10.1145/3735452.3735527},
   booktitle={Proceedings of the 26th ACM SIGPLAN/SIGBED International Conference on Languages, Compilers, and Tools for Embedded Systems},
   publisher={ACM},
   author={Kumar, Abhinav and Khedkar, Atharva and So, Hwisoo and Kuo, Megan and Gurjar, Ameya and Biswas, Partha and Shrivastava, Aviral},
   year={2025},
   month=jun, pages={146–157},
   collection={LCTES ’25} }

@misc{onnx-mlir,
      title={Compiling ONNX Neural Network Models Using MLIR}, 
      author={Tian Jin and Gheorghe-Teodor Bercea and Tung D. Le and Tong Chen and Gong Su and Haruki Imai and Yasushi Negishi and Anh Leu and Kevin O'Brien and Kiyokuni Kawachiya and Alexandre E. Eichenberger},
      year={2020},
      eprint={2008.08272},
      archivePrefix={arXiv},
      primaryClass={cs.PL},
      url={https://arxiv.org/abs/2008.08272}, 
}

@misc{quantum-mlir,
      title={A MLIR Dialect for Quantum Assembly Languages}, 
      author={Alexander McCaskey and Thien Nguyen},
      year={2021},
      eprint={2101.11365},
      archivePrefix={arXiv},
      primaryClass={quant-ph},
      url={https://arxiv.org/abs/2101.11365}, 
}

@misc{suo2025desildetectingsilentbugs,
      title={DESIL: Detecting Silent Bugs in MLIR Compiler Infrastructure}, 
      author={Chenyao Suo and Jianrong Wang and Yongjia Wang and Jiajun Jiang and QingChao Shen and Junjie Chen},
      year={2025},
      eprint={2504.01379},
      archivePrefix={arXiv},
      primaryClass={cs.SE},
      url={https://arxiv.org/abs/2504.01379}, 
}

@misc{GuidanceAI2023,
  author = {{Guidance AI}},
  title = {Guidance: A language model programming framework},
  year = {2023},
  howpublished = {\url{https://github.com/guidance-ai/guidance}},
  note = {Accessed: 2024-07-01}
}

@misc{emitc-tests,
  author = {LLVM Project},
  title = {EmitC Dialect Test Directory},
  year = {2024},
  url = {https://github.com/llvm/llvm-project/tree/main/mlir/test/Dialect/EmitC},
  note = {Accessed: 2024-08-01}
}

@inproceedings{even2023grayc,
  title={Grayc: Greybox fuzzing of compilers and analysers for c},
  author={Even-Mendoza, Karine and Sharma, Arindam and Donaldson, Alastair F and Cadar, Cristian},
  booktitle={Proceedings of the 32nd ACM SIGSOFT International Symposium on Software Testing and Analysis},
  pages={1219--1231},
  year={2023}
}

@inproceedings{jitfuzz,
author = {Wu, Mingyuan and Lu, Minghai and Cui, Heming and Chen, Junjie and Zhang, Yuqun and Zhang, Lingming},
title = {JITfuzz: Coverage-Guided Fuzzing for JVM Just-in-Time Compilers},
year = {2023},
isbn = {9781665457019},
publisher = {IEEE Press},
url = {https://doi.org/10.1109/ICSE48619.2023.00017},
doi = {10.1109/ICSE48619.2023.00017},
booktitle = {Proceedings of the 45th International Conference on Software Engineering},
pages = {56–68},
numpages = {13},
location = {Melbourne, Victoria, Australia},
series = {ICSE '23}
}

@misc{parkGad,
      title={Grammar-Aligned Decoding}, 
      author={Kanghee Park and Jiayu Wang and Taylor Berg-Kirkpatrick and Nadia Polikarpova and Loris D'Antoni},
      year={2024},
      eprint={2405.21047},
      archivePrefix={arXiv},
      primaryClass={cs.AI},
      url={https://arxiv.org/abs/2405.21047}, 
}

@misc{mcmceasy,
      title={Constrained Sampling for Language Models Should Be Easy: An MCMC Perspective}, 
      author={Emmanuel Anaya Gonzalez and Sairam Vaidya and Kanghee Park and Ruyi Ji and Taylor Berg-Kirkpatrick and Loris D'Antoni},
      year={2025},
      eprint={2506.05754},
      archivePrefix={arXiv},
      primaryClass={cs.AI},
      url={https://arxiv.org/abs/2506.05754}, 
}

@misc{syncode,
      title={SynCode: LLM Generation with Grammar Augmentation}, 
      author={Shubham Ugare and Tarun Suresh and Hangoo Kang and Sasa Misailovic and Gagandeep Singh},
      year={2024},
      eprint={2403.01632},
      archivePrefix={arXiv},
      primaryClass={cs.LG},
      url={https://arxiv.org/abs/2403.01632}, 
}

@misc{awrs,
      title={Fast Controlled Generation from Language Models with Adaptive Weighted Rejection Sampling}, 
      author={Benjamin Lipkin and Benjamin LeBrun and Jacob Hoover Vigly and João Loula and David R. MacIver and Li Du and Jason Eisner and Ryan Cotterell and Vikash Mansinghka and Timothy J. O'Donnell and Alexander K. Lew and Tim Vieira},
      year={2025},
      eprint={2504.05410},
      archivePrefix={arXiv},
      primaryClass={cs.CL},
      url={https://arxiv.org/abs/2504.05410}, 
}

@misc{cars,
      title={Constrained Adaptive Rejection Sampling}, 
      author={Paweł Parys and Sairam Vaidya and Taylor Berg-Kirkpatrick and Loris D'Antoni},
      year={2025},
      eprint={2510.01902},
      archivePrefix={arXiv},
      primaryClass={cs.AI},
      url={https://arxiv.org/abs/2510.01902}, 
}

@misc{vllm,
      title={Efficient Memory Management for Large Language Model Serving with PagedAttention}, 
      author={Woosuk Kwon and Zhuohan Li and Siyuan Zhuang and Ying Sheng and Lianmin Zheng and Cody Hao Yu and Joseph E. Gonzalez and Hao Zhang and Ion Stoica},
      year={2023},
      eprint={2309.06180},
      archivePrefix={arXiv},
      primaryClass={cs.LG},
      url={https://arxiv.org/abs/2309.06180}, 
}

@inproceedings {AFLpp,
author = {Andrea Fioraldi and Dominik Maier and Heiko Ei{\ss}feldt and Marc Heuse},
title = {{AFL++} : Combining Incremental Steps of Fuzzing Research},
booktitle = {14th USENIX Workshop on Offensive Technologies (WOOT 20)},
year = {2020},
url = {https://www.usenix.org/conference/woot20/presentation/fioraldi},
publisher = {USENIX Association},
month = aug
}

@ARTICLE{AFLFast,
  author={Böhme, Marcel and Pham, Van-Thuan and Roychoudhury, Abhik},
  journal={IEEE Transactions on Software Engineering}, 
  title={Coverage-Based Greybox Fuzzing as Markov Chain}, 
  year={2019},
  volume={45},
  number={5},
  pages={489-506},
  doi={10.1109/TSE.2017.2785841}}

@inproceedings{FairFuzz,
author = {Lemieux, Caroline and Sen, Koushik},
title = {FairFuzz: a targeted mutation strategy for increasing greybox fuzz testing coverage},
year = {2018},
isbn = {9781450359375},
publisher = {Association for Computing Machinery},
address = {New York, NY, USA},
url = {https://doi.org/10.1145/3238147.3238176},
doi = {10.1145/3238147.3238176},
booktitle = {Proceedings of the 33rd ACM/IEEE International Conference on Automated Software Engineering},
pages = {475–485},
numpages = {11},
location = {Montpellier, France},
series = {ASE '18}
}

@inproceedings {GreyOne,
author = {Shuitao Gan and Chao Zhang and Peng Chen and Bodong Zhao and Xiaojun Qin and Dong Wu and Zuoning Chen},
title = {{GREYONE}: Data Flow Sensitive Fuzzing},
booktitle = {29th USENIX Security Symposium (USENIX Security 20)},
year = {2020},
isbn = {978-1-939133-17-5},
pages = {2577--2594},
url = {https://www.usenix.org/conference/usenixsecurity20/presentation/gan},
publisher = {USENIX Association},
month = aug
}

@inproceedings {EnFuzz,
author = {Yuanliang Chen and Yu Jiang and Fuchen Ma and Jie Liang and Mingzhe Wang and Chijin Zhou and Xun Jiao and Zhuo Su},
title = {{EnFuzz}: Ensemble Fuzzing with Seed Synchronization among Diverse Fuzzers},
booktitle = {28th USENIX Security Symposium (USENIX Security 19)},
year = {2019},
isbn = {978-1-939133-06-9},
address = {Santa Clara, CA},
pages = {1967--1983},
url = {https://www.usenix.org/conference/usenixsecurity19/presentation/chen-yuanliang},
publisher = {USENIX Association},
month = aug
}

@inproceedings {LangFuzz,
author = {Christian Holler and Kim Herzig and Andreas Zeller},
title = {Fuzzing with Code Fragments},
booktitle = {21st USENIX Security Symposium (USENIX Security 12)},
year = {2012},
isbn = {978-931971-95-9},
address = {Bellevue, WA},
pages = {445--458},
url = {https://www.usenix.org/conference/usenixsecurity12/technical-sessions/presentation/holler},
publisher = {USENIX Association},
month = aug
}

@INPROCEEDINGS{SkyFire,
  author={Wang, Junjie and Chen, Bihuan and Wei, Lei and Liu, Yang},
  booktitle={2017 IEEE Symposium on Security and Privacy (SP)}, 
  title={Skyfire: Data-Driven Seed Generation for Fuzzing}, 
  year={2017},
  volume={},
  number={},
  pages={579-594},
  doi={10.1109/SP.2017.23}}

@article{Le12EMI,
author = {Le, Vu and Afshari, Mehrdad and Su, Zhendong},
title = {Compiler validation via equivalence modulo inputs},
year = {2014},
issue_date = {June 2014},
publisher = {Association for Computing Machinery},
address = {New York, NY, USA},
volume = {49},
number = {6},
issn = {0362-1340},
url = {https://doi.org/10.1145/2666356.2594334},
doi = {10.1145/2666356.2594334},
journal = {SIGPLAN Not.},
month = jun,
pages = {216–226},
numpages = {11}
}

@inproceedings{alive,
author = {Lopes, Nuno P. and Menendez, David and Nagarakatte, Santosh and Regehr, John},
title = {Provably correct peephole optimizations with alive},
year = {2015},
isbn = {9781450334686},
publisher = {Association for Computing Machinery},
address = {New York, NY, USA},
url = {https://doi.org/10.1145/2737924.2737965},
doi = {10.1145/2737924.2737965},
booktitle = {Proceedings of the 36th ACM SIGPLAN Conference on Programming Language Design and Implementation},
pages = {22–32},
numpages = {11},
location = {Portland, OR, USA},
series = {PLDI '15}
}

@inproceedings{alive2,
author = {Lopes, Nuno P. and Lee, Juneyoung and Hur, Chung-Kil and Liu, Zhengyang and Regehr, John},
title = {Alive2: bounded translation validation for LLVM},
year = {2021},
isbn = {9781450383912},
publisher = {Association for Computing Machinery},
address = {New York, NY, USA},
url = {https://doi.org/10.1145/3453483.3454030},
doi = {10.1145/3453483.3454030},
booktitle = {Proceedings of the 42nd ACM SIGPLAN International Conference on Programming Language Design and Implementation},
pages = {65–79},
numpages = {15},
location = {Virtual, Canada},
series = {PLDI 2021}
}

@misc{Souper,
      title={Souper: A Synthesizing Superoptimizer}, 
      author={Raimondas Sasnauskas and Yang Chen and Peter Collingbourne and Jeroen Ketema and Gratian Lup and Jubi Taneja and John Regehr},
      year={2018},
      eprint={1711.04422},
      archivePrefix={arXiv},
      primaryClass={cs.PL},
      url={https://arxiv.org/abs/1711.04422}, 
}

@article{MLIRTracer,
  title={Directed Testing in MLIR: Unleashing Its Potential by Overcoming the Limitations of Random Fuzzing},
  author={Tong, Weiyuan and Wang, Zixu and Tang, Zhanyong and Fang, Jianbin and Zhang, Yuqun and Ye, Guixin},
  journal={Proceedings of the ACM on Software Engineering},
  volume={2},
  number={FSE},
  pages={2288--2310},
  year={2025},
  publisher={ACM New York, NY, USA}
}

@inproceedings{ZellerLearnGrammar,
author = {Gopinath, Rahul and Mathis, Bj\"{o}rn and Zeller, Andreas},
title = {Mining input grammars from dynamic control flow},
year = {2020},
isbn = {9781450370431},
publisher = {Association for Computing Machinery},
address = {New York, NY, USA},
url = {https://doi.org/10.1145/3368089.3409679},
doi = {10.1145/3368089.3409679},
booktitle = {Proceedings of the 28th ACM Joint Meeting on European Software Engineering Conference and Symposium on the Foundations of Software Engineering},
pages = {172–183},
numpages = {12},
location = {Virtual Event, USA},
series = {ESEC/FSE 2020}
}

@inproceedings{Polyglot,
  title={One engine to fuzz’em all: Generic language processor testing with semantic validation},
  author={Chen, Yongheng and Zhong, Rui and Hu, Hong and Zhang, Hangfan and Yang, Yupeng and Wu, Dinghao and Lee, Wenke},
  booktitle={2021 IEEE Symposium on Security and Privacy (SP)},
  pages={642--658},
  year={2021},
  organization={IEEE}
}

@article{LLMTest,
   title={LLM-Based Test-Driven Interactive Code Generation: User Study and Empirical Evaluation},
   volume={50},
   ISSN={2326-3881},
   url={http://dx.doi.org/10.1109/TSE.2024.3428972},
   DOI={10.1109/tse.2024.3428972},
   number={9},
   journal={IEEE Transactions on Software Engineering},
   publisher={Institute of Electrical and Electronics Engineers (IEEE)},
   author={Fakhoury, Sarah and Naik, Aaditya and Sakkas, Georgios and Chakraborty, Saikat and Lahiri, Shuvendu K.},
   year={2024},
   month=sep, pages={2254–2268} }

@misc{NeuralGuided,
      title={Neural Program Synthesis with a Differentiable Fixer}, 
      author={Matej Balog and Rishabh Singh and Petros Maniatis and Charles Sutton},
      year={2020},
      eprint={2006.10924},
      archivePrefix={arXiv},
      primaryClass={stat.ML},
      url={https://arxiv.org/abs/2006.10924}, 
}

@misc{LearnFuzz,
      title={Learn\&Fuzz: Machine Learning for Input Fuzzing}, 
      author={Patrice Godefroid and Hila Peleg and Rishabh Singh},
      year={2017},
      eprint={1701.07232},
      archivePrefix={arXiv},
      primaryClass={cs.AI},
      url={https://arxiv.org/abs/1701.07232}, 
}

@misc{LLMDLFuzz,
      title={Large Language Models are Zero-Shot Fuzzers: Fuzzing Deep-Learning Libraries via Large Language Models}, 
      author={Yinlin Deng and Chunqiu Steven Xia and Haoran Peng and Chenyuan Yang and Lingming Zhang},
      year={2023},
      eprint={2212.14834},
      archivePrefix={arXiv},
      primaryClass={cs.SE},
      url={https://arxiv.org/abs/2212.14834}, 
}

\newpage
\appendix
\section{Hardware and Software}
\label{app:hardware} 
Our experiments were conducted on Ubuntu 22.04 LTS nodes with Intel Xeon Gold 6230 CPUs (2.10 GHz, 10 cores, 20 threads allocated) and 384 GB RAM. For GPU-accelerated workloads, we provisioned 1x NVIDIA RTX A6000 GPUs. Our implementation is based on Python 3.10.12, PyTorch 2.8.0 with CUDA 12.8, Transformers 4.55.4 and llguidance 0.7.30.

\section{Hyperparameters}
\label{app:hyperparams}
For the language model decoding, we set the temperature to 1.0, top-p to 1.0, and top-k to 0 to allow sampling from the full token vocabulary without distributional distortion. We set the maximum number of newly generated tokens to 1024 tokens.
\section{Model Checkpoint}
\label{app:models}
We evaluate on one instruction-tuned model:
\begin{itemize}
    \item \textbf{Qwen/Qwen2.5-Coder-7B-Instruct}~\citep{qwen}: \url{https://huggingface.co/Qwen/Qwen2.5-Coder-7B-Instruct} (commit \texttt{a09a354})
\end{itemize}
The model uses BF16 precision with its default tokenizer and system prompt.

\section{System Overview}
\label{app:overview}

\begin{figure}[h]
\centering
\scalebox{0.7}{
\begin{tikzpicture}[node distance=0.6cm and 1.2cm, font=\sffamily]

\node[data] (sut) {System Under Test};

\node[process, below=0.4cm of sut] (extract) {Corpus \& Specifications Extraction};

\node[data, below left=0.5cm and 0.3cm of extract] (testcases) {Test Cases $\mathcal{E}$};
\node[data, below right=0.5cm and 0.3cm of extract] (tablegen) {TableGen Files $\mathcal{T}$};

\draw[arrow] (sut) -- (extract);
\draw[arrow] (extract) -- (testcases);
\draw[arrow] (extract) -- (tablegen);

Background box for Phase 1
\begin{scope}[on background layer]
\node[component, fill=bg1, fit=(sut)(extract)(testcases)(tablegen), inner sep=14pt, label={[anchor=north west, font=\footnotesize\sffamily\bfseries]north west:Phase 1: Extraction}] (phase1) {};
\end{scope}

\node[process, below=2.8cm of extract] (parse) {Parse TableGen Specs};
\node[data, below=0.5cm of parse] (explicit) {Explicit Formats};
\node[data, right=1.8cm of explicit] (custom) {Custom Formats};

\draw[arrow] (tablegen) -- (parse);
\draw[arrow] (parse) -- (explicit);
\draw[arrow] (parse) -- (custom);

\node[process, below=1.1cm of explicit] (translate) {Format $\rightarrow$ Lark rules};
\draw[arrow] (explicit) -- (translate);

\node[process, below=0.5cm of custom] (match) {Match Op with Examples};
\node[decision, below right=0.2cm and 0.4cm of match] (found) {Examples Found?};

\draw[arrow] (custom) -- (match);
\draw[arrow] (testcases.south) .. controls +(0,-4.5) and +(-2,0) .. (match.west);
\draw[arrow] (match) -- (found);

\node[process, below left=1.1cm and 0.3cm of found] (infer) {LM Inference + Verify};
\node[process, below right=1.1cm and 0.3cm of found] (generic) {Mark as Generic};

\draw[arrow] (found) -- node[left, font=\footnotesize, pos=0.15] {yes} (infer);
\draw[arrow] (found) -- node[right, font=\footnotesize, pos=0.15] {no} (generic);

\node[data, below=0.7cm of translate, fill=tertiary!20] (grammar) {Dialect Grammar $\mathcal{G}_{\mathcal{D}}$};

\draw[arrow] (translate) -- (grammar);
\draw[arrow] (infer) |- (grammar);
\draw[arrow] (generic) |- (grammar);

\begin{scope}[on background layer]
\node[component, fill=bg2, fit=(parse)(explicit)(custom)(translate)(match)(found)(infer)(generic)(grammar), inner sep=18pt,
      label={[anchor=north west, font=\footnotesize\sffamily\bfseries]north west:Phase 2: Grammar Extraction (\S\ref{sec:grammar-extraction})}] (phase2) {};
\end{scope}

\node[process, below=2.8cm of grammar] (gcd) {Grammar-Constrained Sampling};
\node[data, left=1.5cm of gcd] (lm) {Pre-trained LM $\mathcal{M}$};

\draw[arrow] (grammar) -- (gcd);
\draw[arrow] (lm) -- (gcd);
\draw[arrow] (testcases.south) .. controls +(0,-2.5) and +(-3.5,0) .. (gcd.west) node[pos=0.7, left, font=\footnotesize, text width=1.5cm, align=center] {few-shot prompts};

\node[data, below=0.5cm of gcd] (asm) {Seeds in Assembly (or fallback Generic)};
\draw[arrow] (gcd) -- (asm);

\node[process, below=0.5cm of asm] (convert) {\texttt{mlir-print-op-generic}};
\draw[arrow] (asm) -- (convert);

\node[data, below=0.5cm of convert, fill=tertiary!20] (seeds) {Seed Corpus $\mathcal{S}$ (Generic Format)};
\draw[arrow] (convert) -- (seeds);

\begin{scope}[on background layer]
\node[component, fill=bg3, fit=(lm)(gcd)(asm)(convert)(seeds), inner sep = 18pt,
      label={[anchor=north west, font=\footnotesize\sffamily\bfseries]north west:Phase 3: Seed Generation (\S\ref{sec:gcd-implementation})}] (phase3) {};
\end{scope}

\node[process, right=2.4cm of seeds, fill=secondary!20] (synthfuzz) {SynthFuzz Mutation Engine};
\node[process, below=0.5cm of synthfuzz] (execute) {Execute on SUT};
\node[data, below=0.5cm of execute] (coverage) {Coverage Feedback};

\draw[bigarrow] (seeds) -- (synthfuzz);
\draw[arrow] (synthfuzz) -- (execute);
\draw[arrow] (execute) -- (coverage);

\draw[arrow, dashed, line width=1pt] (coverage.west) .. controls +(-1.5,0) and +(-1.5,0) .. (synthfuzz.west) 
    node[pos=0.5, left, font=\footnotesize] {adapt};

\node[data, right=1.2cm of execute, fill=secondary!30] (bugs) {\textbf{Bugs}};
\draw[arrow, secondary, thick] (execute) -- (bugs);

\begin{scope}[on background layer]
\node[component, fill=gray!10, fit=(synthfuzz)(execute)(coverage)(bugs), inner sep = 18pt, 
      label={[anchor=north west, font=\footnotesize\sffamily\bfseries]north west:Phase 4: Coverage-Guided Fuzzing (\S\ref{sec:fuzzing-loop})}] (phase4) {};
\end{scope}


\node[below=0.1cm of synthfuzz.east, anchor=west, font=\footnotesize, text width=3cm, align=left] 
    {\textit{SynthFuzz is dialect-agnostic but corpus-dependent. Phases 1--3 enable bootstrapping.}};
\draw[decorate, decoration={brace, amplitude=5pt, mirror}] 
    ([xshift=-2pt]asm.south west) -- ([xshift=-6pt]seeds.south west) 
    node[midway, left=5pt, font=\footnotesize, text width=2.2cm, align=right] {Assembly $\rightarrow$ Generic\\(exposes bugs)};

\end{tikzpicture}
}
\caption{Accuracy vs. efficiency for different sampling methods.}
\label{fig:germinator-overview}
\end{figure}
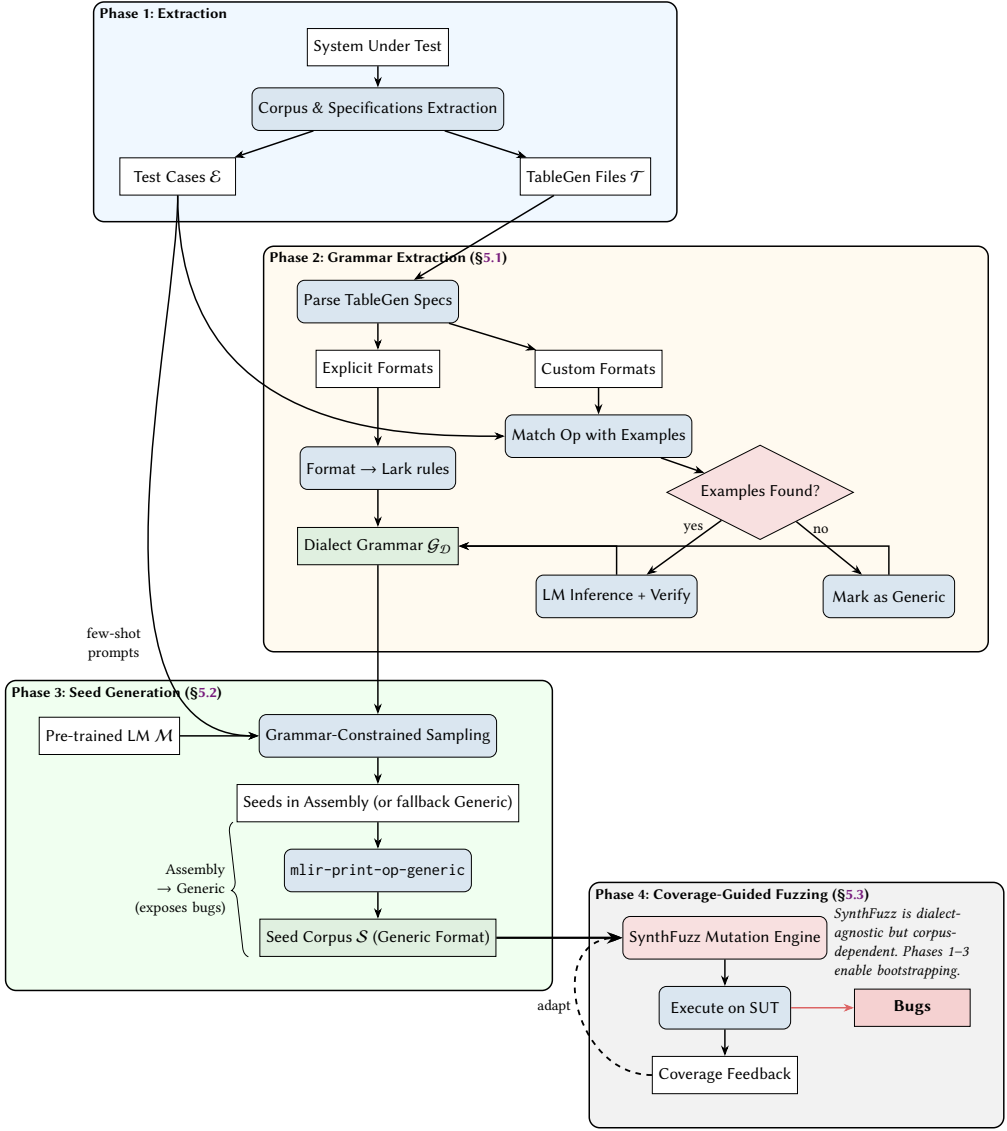

Figure~\ref{fig:germinator-overview} provides a high-level overview of \name's architecture and workflow.
The system takes as input a dialect specification in TableGen and produces valid seed programs through three main stages:
(1) grammar extraction from TableGen specifications (\Cref{sec:grammar-extraction}),
(2) LM-guided seed generation with grammar-constrained decoding (\Cref{sec:gcd-implementation}), and
(3) coverage-guided fuzzing using the generated seeds (\Cref{sec:fuzzing-loop}).
The extracted grammars serve as constraints during LM decoding, ensuring syntactic validity by construction, while the LM's learned priors guide generation toward semantically plausible programs.

\section{Evaluation Target Projects}
\label{app:targets}

We selected six projects spanning diverse maturity levels and application domains to evaluate \name's effectiveness across the MLIR ecosystem (Table~\ref{tab:targets}).
Our selection criteria ensure representation of:
\textbf{(1) Mature ecosystems} with extensive test infrastructure (MLIR with 2,000 tests, IREE with 1,100 tests),
\textbf{(2) High-profile projects} with limited testing despite significant community adoption (Triton: 17,500 GitHub stars but only 80 tests), and
\textbf{(3) Emerging dialects} where test creation lags behind development (HEIR: 600 stars, 170 tests).

This diversity demonstrates \name's effectiveness across different resource settings: from well-tested infrastructure dialects to under-tested but widely-adopted production systems.

\begin{table}[h]
\caption{Target projects with diverse maturity levels. Despite high community adoption, projects like Triton maintain limited test infrastructure, highlighting the need for automated test generation.}
\small
\centering
\begin{tabular}{lrrr}
\toprule
\textbf{Project} & \textbf{Domain} & \textbf{Stars} & \textbf{Tests} \\
\midrule
MLIR & Infrastructure & 35,200 & 2,000 \\
Torch-MLIR & ML Frontend & 1,700 & 60 \\
IREE & ML Runtime & 3,400 & 1,100 \\
CIRCT & Hardware & 1,900 & 650 \\
Triton & GPU Kernels & 17,500 & 80 \\
HEIR & Cryptography & 600 & 170 \\
\bottomrule
\end{tabular}
\label{tab:targets}
\end{table}
\section{LLM-Based Semantic Filtering}
\label{app:llm-judge}

When syntactically valid programs fail the MLIR verifier's semantic checks, \name faces a filtering decision: which failures represent "nearly correct" programs (likely repairable through mutation) versus fundamentally malformed ones?

\subsection{Why Distance Metrics Fail}

Traditional distance metrics (e.g., edit distance, AST similarity) are inadequate for judging semantic proximity in MLIR:
\begin{itemize}
    \item \textbf{Small syntactic changes cause large semantic differences:} Swapping SSA value names (\texttt{\%0} $\leftrightarrow$ \texttt{\%1}) is a minimal edit but can violate def-use chains.
    \item \textbf{Large syntactic differences may be semantically equivalent:} Operation reordering in commutative contexts involves significant structural changes but preserves semantics.
    \item \textbf{Semantic violations span heterogeneous categories:} Type errors, SSA violations, and dominance violations cannot be meaningfully compared via edit distance.
\end{itemize}

\subsection{LLM-Based Plausibility Scoring}

Instead, \name queries the LM to score semantic plausibility. For each syntactically valid but semantically invalid program $p$, we prompt the LM:

\begin{quote}
\texttt{Rate the semantic plausibility ($\tau$) of this MLIR program on a scale of 0-100, where 100 is perfectly valid and 0 is nonsensical. Consider SSA form, type consistency, and common MLIR patterns. Output only a number.}
\end{quote}

Programs scoring above threshold $\tau=70$ are retained as seeds. The LM's exposure to MLIR code during pre-training enables it to distinguish realistic semantic patterns from fundamentally malformed structures.

\subsection{Implementation Details}

We use the same LM (Qwen/Qwen2.5-Coder-7B-Instruct) for both generation and judging to avoid additional model overhead ($\sim$200ms per program).
\section{Grammar Extraction Statistics}
\label{app:grammar-extraction}

Table~\ref{tab:grammar-extraction} reports detailed grammar extraction success rates across all dialects in our six evaluated projects.

\begin{table}[t]
\caption{Grammar extraction success rates by automation level. \textit{Automatic}: Extracted directly from TableGen. \textit{Inferred}: Derived from test examples. \textit{Failed}: Cannot extract due to missing tests and custom specs.}
\small
\centering
\begin{tabular}{lrrrrr}
\toprule
\textbf{Project} & \textbf{Total} & \textbf{Total} & \textbf{Automatic} & \textbf{Inferred} & \textbf{Failed} \\
 & \textbf{Dialects} & \textbf{Ops} & \textbf{(\%)} & \textbf{(\%)} & \textbf{(\%)} \\
\midrule
MLIR & 45 & 964 & 83.5\% & 7.0\% & 9.5\% \\
Torch-MLIR & 7 & 215 & 56.0\% & 20.0\% & 24.0\% \\
IREE & 38 & 659 & 91.0\% & 1.5\% & 7.5\% \\
CIRCT & 20 & 114 & 87.0\% & 3.0\% & 10.0\% \\
Triton & 13 & 76 & 86.0\% & 5.0\% & 9.0\% \\
HEIR & 14 & 75 & 93.0\% & 0.0\% & 7.0\% \\
\midrule
\textbf{Average} & --- & --- & \textbf{82.8\%} & \textbf{6.1\%} & \textbf{11.1\%} \\
\bottomrule
\end{tabular}
\label{tab:grammar-extraction}
\end{table}
\section{Detailed Bug Discovery Statistics}
\label{app:bug-counts}

Table~\ref{tab:bug-summary} provides complete bug counts for all six projects after 24-hour fuzzing campaigns.

\begin{table}[t]
\caption{Bug counts after 24-hour campaigns. Improvement calculated against the best baseline. Where baselines found zero bugs, improvement marked as ``--'' to indicate qualitative advantage.}
\small
\centering
\begin{tabular}{lrrrr}
\toprule
\textbf{Project} & \textbf{Grammarinator} & \textbf{SynthFuzz} & \textbf{\name} & \textbf{Improvement} \\
\midrule
MLIR & 3 & 6 & 25 & 4.13$\times$ \\
Torch-MLIR & 0 & 0 & 7 & -- \\
IREE & 3 & 5 & 25 & 5.0$\times$ \\
CIRCT & 1 & 3 & 15 & 5.0$\times$ \\
Triton & 0 & 0 & 3 & -- \\
HEIR & 1 & 2 & 13 & 6.5$\times$ \\
\midrule
\textbf{Total} & \textbf{8} & \textbf{16} & \textbf{88} & \textbf{5.5$\times$} \\
\bottomrule
\end{tabular}
\label{tab:bug-summary}
\end{table}
\section{Prompt Templates}
\label{app:prompts}

\name automatically constructs prompts for seed generation without manual curation. This section provides representative templates for both assembly and generic format generation, as well as few-shot and zero-shot scenarios.

\subsection{Assembly Format Prompt Template}

When grammar extraction successfully produces assembly format rules, \name constructs prompts using the following template:

\begin{lstlisting}[basicstyle=\footnotesize\ttfamily, breaklines=true, numbers=none]
You are an MLIR code generator for the <DIALECT_NAME> dialect.

Generate syntactically valid MLIR programs that:
1. Use only operations from: <SUPPORTED_OPS_LIST>
2. Follow MLIR semantic constraints (SSA form, type consistency, dominance)
3. Explore diverse operation combinations and control flow patterns

Target Operations (prioritized by rarity in test corpus):
<TOP_K_RARE_OPS_WITH_USAGE_COUNTS>

Complexity Target: ~<AVG_OPS> operations per program

Grammar Constraints:
<EXTRACTED_GRAMMAR_RULES>

Few-Shot Examples:
<EXAMPLE_1>
<EXAMPLE_2>
...
<EXAMPLE_N>

Generate ONE novel MLIR program following the grammar and semantic rules.
Output only raw MLIR code.
\end{lstlisting}

\textbf{Placeholders:}
\begin{itemize}[noitemsep]
\item \texttt{<DIALECT\_NAME>}: Target dialect (e.g., \texttt{torch}, \texttt{emitc})
\item \texttt{<SUPPORTED\_OPS\_LIST>}: Operations with successfully extracted grammars
\item \texttt{<TOP\_K\_RARE\_OPS>}: Operations ranked by inverse frequency in test corpus
\item \texttt{<AVG\_OPS>}: Mean operation count from existing tests (or default 6-8)
\item \texttt{<EXTRACTED\_GRAMMAR\_RULES>}: Lark grammar rules from TableGen extraction
\item \texttt{<EXAMPLE\_1...N>}: 3-5 examples selected by structural diversity (AST depth, operation variety)
\end{itemize}

All placeholders are filled automatically from dialect specifications and test corpora.

\subsection{Generic Format Prompt Template}

When grammar extraction fails or falls back to generic format:

\begin{lstlisting}[basicstyle=\footnotesize\ttfamily, breaklines=true, numbers=none]
You are an MLIR code generator for the <DIALECT_NAME> dialect.

Generate syntactically valid MLIR programs in GENERIC FORMAT:
- Syntax: "dialect.op"(operands) {regions} attributes : (types) -> (types)
- Use only operations from: <AVAILABLE_OPS>
- Follow SSA form, type consistency, and dominance rules

Available Operations:
<OP_NAME_1>: <SIGNATURE_1>
<OP_NAME_2>: <SIGNATURE_2>
...

Common Type Vocabularies:
<TYPE_CONSTRAINTS>

Target: ~<AVG_OPS> operations, <REGION_DEPTH> region nesting

<IF_FEW_SHOT>
Few-Shot Examples:
<EXAMPLE_1>
...
<EXAMPLE_N>
</IF_FEW_SHOT>

<IF_ZERO_SHOT>
Structural Scaffolds:
- Basic module: "builtin.module"() ({...}) : () -> ()
- Function: "func.func"() <{...}> ({...}) : () -> ()
- Region syntax: ({^bb0: operations})
</IF_ZERO_SHOT>

Generate ONE novel generic-format MLIR program.
Output only raw MLIR code.
\end{lstlisting}

\textbf{Placeholders:}
\begin{itemize}[noitemsep]
\item \texttt{<AVAILABLE\_OPS>}: All operations in dialect (from TableGen)
\item \texttt{<SIGNATURE\_X>}: Operation signatures extracted from TableGen (operand/result types)
\item \texttt{<TYPE\_CONSTRAINTS>}: Valid type vocabularies (e.g., \texttt{IntegerIndexOrOpaqueType})
\item \texttt{<IF\_FEW\_SHOT> / <IF\_ZERO\_SHOT>}: Conditional sections based on test availability
\item \texttt{<REGION\_DEPTH>}: Target nesting level (default 2-4)
\end{itemize}

\subsection{Zero-Shot Prompt (Low-Resource Dialects)}

For dialects with no existing tests:

\begin{lstlisting}[basicstyle=\footnotesize\ttfamily, breaklines=true, numbers=none]
You are an MLIR code generator for the <DIALECT_NAME> dialect.

This dialect has no existing test corpus. Generate valid MLIR programs using:

Available Operations (from TableGen):
<OP_NAME>(<OPERAND_TYPES>) -> <RESULT_TYPES>
...

MLIR Structural Requirements:
1. SSA form: Each value defined once, used after definition
2. Modules contain functions: "builtin.module"() ({...})
3. Functions have signatures: "func.func"() <{function_type = ...}>
4. Regions need terminators: "func.return"(...) at function end

Syntactic Templates:
- Generic op: "<dialect>.<op>"(%arg1, %arg2) : (type1, type2) -> type3
- Region: ({^bb0(%arg: type): <operations> })
- Attribute: <{key = value}>

Generate ONE valid MLIR program exploring <DIALECT_NAME> operations.
\end{lstlisting}

\subsection{Prompt Construction Algorithm}

\begin{algorithm}
\caption{Automatic Prompt Construction}
\small
\DontPrintSemicolon
\KwIn{Dialect $\mathcal{D}$, extracted grammar $G_{\mathcal{D}}$, test corpus $T$}
\KwOut{Prompt string $P$}

$P \gets \text{BaseTemplate}(\mathcal{D})$\;
\tcp{Select format based on grammar extraction}
\eIf{$G_{\mathcal{D}}$ is assembly format}{
    $P \gets P + \text{AssemblyGrammarRules}(G_{\mathcal{D}})$\;
}{
    $P \gets P + \text{GenericFormatTemplate}()$\;
    $P \gets P + \text{OperationSignatures}(\mathcal{D})$\;
}

\tcp{Add examples if available}
\eIf{$|T| > 0$}{
    $examples \gets \text{SelectDiverse}(T, k=4)$ \tcp*{By AST depth, op variety}
    $P \gets P + \text{FewShotExamples}(examples)$\;
}{
    $P \gets P + \text{ZeroShotScaffolds}()$\;
}

\tcp{Add operational guidance}
$rare\_ops \gets \text{RankByRarity}(\mathcal{D}, T)$\;
$P \gets P + \text{RareOperationList}(rare\_ops[:10])$\;
$P \gets P + \text{ComplexityTargets}(\text{Mean}(|T|), \text{AvgDepth}(T))$\;

\Return{$P$}\;
\end{algorithm}

All prompt construction is fully automated, requiring no manual intervention per dialect.
\section{Base MLIR Grammar Template}
\label{app:grammar-template}

\name extends a base MLIR grammar template with dialect-specific rules extracted from TableGen specifications.
This section presents the core grammar template in Lark format, highlighting extension points where dialect-specific productions are inserted.

\subsection{Grammar Structure}

The base grammar provides the foundation for all MLIR programs, including:
\begin{itemize}[noitemsep]
\item \textbf{Type system}: Standard types (integers, floats, tensors, memrefs), function types, and dialect type extension points
\item \textbf{Attribute system}: Standard attributes (integers, floats, strings, arrays, dictionaries) and dialect attribute extension points
\item \textbf{SSA infrastructure}: SSA value definitions, uses, and type annotations
\item \textbf{Control flow}: Blocks, regions, successors, and branch targets
\item \textbf{Module structure}: Top-level modules, functions, and nested regions
\item \textbf{Affine maps}: Expressions, constraints, and integer sets for polyhedral compilation
\end{itemize}

\subsection{Extension Points}

The template contains three primary extension points marked with placeholder tags:

\begin{enumerate}[noitemsep]
\item \textbf{\texttt{\{\{OPERATION\_RULES\}\}}}: Individual operation grammar rules extracted from TableGen
\item \textbf{\texttt{\{\{DIALECT\_AGGREGATES\}\}}}: Dialect-level aggregation rules grouping related operations
\item \textbf{\texttt{\{\{DIALECT\_LIST\}\}}}: Union of all dialect operation rules
\end{enumerate}

For example, after extracting rules for the EmitC dialect, \name instantiates these placeholders as:

\begin{lstlisting}[basicstyle=\scriptsize\ttfamily, numbers=none]
// {{OPERATION_RULES}} becomes:
emitc_literal: op_result_list? "emitc.literal" 
    ESCAPED_STRING attribute_dict? ":" type
emitc_for: "emitc.for" SSA_ID "=" ssa_use "to" ssa_use 
    "step" ssa_use (":" integer_index_or_opaque)? region
emitc_yield: "emitc.yield" attribute_dict? 
    (ssa_use ":" type)?

// {{DIALECT_AGGREGATES}} becomes:
emitc_ops: emitc_literal | emitc_for | emitc_yield | ...

// {{DIALECT_LIST}} becomes:
all_dialect_ops: emitc_ops | arith_ops | func_ops | ...
\end{lstlisting}

\subsection{Abbreviated Grammar}

The complete base grammar spans approximately 300 lines.
Below is an abbreviated version highlighting key structures and extension points:

\begin{lstlisting}[basicstyle=\tiny\ttfamily, breaklines=true, numbers=none]
start: mlir_file
mlir_file: definition_and_module_list+ | definition_and_function_list+

// ============================================================================
// TYPES
// ============================================================================
type: TYPE_ALIAS | dialect_type | standard_type
standard_type: complex_type | FLOAT_TYPE | function_type | INDEX_TYPE 
             | INTEGER_TYPE | memref_type | tensor_type | ...
dialect_type: "!" (opaque_dialect_item | pretty_dialect_item)

tensor_type: ranked_tensor_type | unranked_tensor_type
ranked_tensor_type: "tensor" "<" dimension_list_ranked? 
                    tensor_memref_element_type ">"
memref_type: ranked_memref_type | unranked_memref_type

// {{CUSTOM_TYPES}} - Placeholder for dialect-specific type vocabularies

// ============================================================================
// ATTRIBUTES
// ============================================================================
attribute_value: ATTRIBUTE_ALIAS | dialect_attribute | standard_attribute
standard_attribute: array_attribute | BOOL | dictionary_attribute 
                  | elements_attribute | float_attribute | ...
dialect_attribute: "#" (opaque_dialect_item | pretty_dialect_item)

// {{CUSTOM_ATTRIBUTES}} - Placeholder for dialect-specific attributes

// ============================================================================
// SSA VALUES AND OPERATIONS
// ============================================================================
ssa_use: SSA_ID | CONSTANT
ssa_use_list: ssa_use ("," ssa_use)*
op_result_list: op_result ("," op_result)* "="

// {{OPERATION_RULES}} - Individual dialect operation grammar rules
// Example: emitc_literal: op_result_list? "emitc.literal" ...

// {{DIALECT_AGGREGATES}} - Dialect-level groupings
// Example: emitc_ops: emitc_literal | emitc_for | emitc_yield

all_dialect_ops: {{DIALECT_LIST}}
operation: (all_dialect_ops | module | function) location_attribute?

// ============================================================================
// REGIONS AND CONTROL FLOW
// ============================================================================
region: "{" block* "}"
block: block_label? operation_list
block_label: BLOCK_ID ("(" ssa_id_and_type_list? ")")? ":"

// ============================================================================
// MODULES AND FUNCTIONS
// ============================================================================
module: "module" SYMBOL_REF_ID? ("attributes" attribute_dict)? region
function: "func.func" SYMBOL_REF_ID "(" argument_list? ")" 
          ("->" function_result_list)? region?

// ============================================================================
// TERMINALS
// ============================================================================
SSA_ID: "%" SUFFIX_ID ("#" DIGITS)?
SYMBOL_REF_ID: "@" (SUFFIX_ID | ESCAPED_STRING)
BLOCK_ID: "^" SUFFIX_ID
TYPE_ALIAS: "!" (ESCAPED_STRING | BARE_ID)
INDEX_TYPE: "index"
INTEGER_TYPE: /si?[1-9][0-9]*/ | /ui[1-9][0-9]*/
FLOAT_TYPE: "f16" | "bf16" | "f32" | "f64"
BARE_ID: /[a-zA-Z_][a-zA-Z0-9_$]*/
\end{lstlisting}

The full grammar template is available in the \name repository at \texttt{grammars/mlir\_base.lark}.

\subsection{Grammar Instantiation Process}

For each dialect $\mathcal{D}$, \name:
\begin{enumerate}[noitemsep]
\item Extracts operation rules from TableGen $\rightarrow$ \texttt{\{\{OPERATION\_RULES\}\}}
\item Groups operations by dialect $\rightarrow$ \texttt{\{\{DIALECT\_AGGREGATES\}\}}
\item Constructs union of all dialects $\rightarrow$ \texttt{\{\{DIALECT\_LIST\}\}}
\item Adds type vocabulary rules (e.g., \texttt{integer\_index\_or\_opaque}) from TableGen constraints
\item Generates final Lark grammar file: \texttt{mlir\_\$\{dialect\}.lark}
\end{enumerate}

This modular design enables efficient grammar reuse across dialects while maintaining precise dialect-specific constraints.
\section{Example Bug Reports}
\label{app:bug-reports}

This section presents representative bugs discovered by \name across different MLIR dialects, demonstrating the tool's ability to find diverse crash and verification failures. Each bug includes a minimal reproducer that triggers the issue.

\subsection{Parser Crash: Mixed Type Literals in Dense Attributes}

\textbf{Dialect:} TOSA \\
\textbf{Component:} Attribute parser \\
\textbf{Symptom:} Assertion failure in \texttt{getIntAttrElements()} when dense tensor attribute contains type-mismatched literals

\textbf{Minimal reproducer (generic format):}
\begin{lstlisting}[language=MLIR, basicstyle=\scriptsize\ttfamily]
"builtin.module"() ({
  "tosa.const"() <{value = dense<["string", 1]> : tensor<2xi32>}> 
    : () -> tensor<2xi32>
}) : () -> ()
\end{lstlisting}

\textbf{Root cause:} The parser assumes all elements in a dense integer tensor are integers or booleans, but does not validate this before attempting to parse. When a string literal appears, \texttt{getIntAttrElements()} asserts on line 642 of \texttt{AttributeParser.cpp}.

\textbf{Expected behavior:} Parser should emit a diagnostic error message rather than crashing.

\subsection{Verifier Crash: Negative Array Attribute Index}

\textbf{Dialect:} GPU \\
\textbf{Component:} Operation verifier \\
\textbf{Symptom:} Heap buffer overflow when accessing negative \texttt{workgroup\_attributions} index

\textbf{Minimal reproducer (assembly format):}
\begin{lstlisting}[language=MLIR, basicstyle=\scriptsize\ttfamily]
gpu.module @memref_conversions {
  gpu.func @kern(%arg0: memref<8xf32>) kernel 
      workgroup(%wg: memref<8xf32>) 
      attributes {workgroup_attributions = -1 : i64} {
    gpu.return
  }
}
\end{lstlisting}

\textbf{Equivalent generic format:}
\begin{lstlisting}[language=MLIR, basicstyle=\scriptsize\ttfamily]
"builtin.module"() ({
  "gpu.module"() <{sym_name = "memref_conversions"}> ({
    "gpu.func"() <{function_type = (memref<8xf32>) -> ()}> ({
    ^bb0(%arg0: memref<8xf32>):
      "gpu.return"() : () -> ()
    }) {gpu.kernel, sym_name = "kern", 
        workgroup_attributions = -1 : i64} : () -> ()
  }) : () -> ()
}) : () -> ()
\end{lstlisting}

\textbf{Root cause:} The verifier's \texttt{verifyAttributions()} function (line 552 of \texttt{GPUDialect.cpp}) accesses block arguments using the \texttt{workgroup\_attributions} value without bounds checking. A negative value causes out-of-bounds access.

\textbf{Expected behavior:} Verifier should reject negative attribution counts with a diagnostic error.

\subsection{Verifier Crash: Missing Loop Induction Variable}

\textbf{Dialect:} EmitC \\
\textbf{Component:} Operation verifier \\
\textbf{Symptom:} Assertion failure when \texttt{emitc.for} loop body has no block arguments

\textbf{Minimal reproducer (assembly format):}
\begin{lstlisting}[language=MLIR, basicstyle=\scriptsize\ttfamily]
func.func @test() {
  %lb = emitc.literal "0" : index
  %ub = emitc.literal "10" : index  
  %step = emitc.literal "1" : index
  emitc.for %lb to %ub step %step : index {
    emitc.yield
  }
  func.return
}
\end{lstlisting}

\textbf{Equivalent generic format:}
\begin{lstlisting}[language=MLIR, basicstyle=\scriptsize\ttfamily]
"builtin.module"() ({
  "func.func"() <{function_type = () -> (), sym_name = "test"}> ({
    %0 = "emitc.literal"() <{value = "0"}> : () -> index
    %1 = "emitc.literal"() <{value = "10"}> : () -> index  
    %2 = "emitc.literal"() <{value = "1"}> : () -> index
    "emitc.for"(%0, %1, %2) ({
      "emitc.yield"() : () -> ()
    }) : (index, index, index) -> ()
    "func.return"() : () -> ()
  }) : () -> ()
}) : () -> ()
\end{lstlisting}

\textbf{Root cause:} The \texttt{getInductionVar()} method assumes the loop region always contains a block argument (the induction variable) and attempts \texttt{getBody()->getArgument(0)} without checking if any arguments exist.

\textbf{Expected behavior:} While the TableGen specification permits loops without explicit induction variables, the verifier should either require them or handle their absence gracefully.

\subsection{Type System Crash: Invalid Tensor Element Type}

\textbf{Dialect:} Torch \\
\textbf{Component:} Type verifier \\
\textbf{Symptom:} Segmentation fault when tensor uses unsupported element type

\textbf{Minimal reproducer (generic format):}
\begin{lstlisting}[language=MLIR, basicstyle=\scriptsize\ttfamily]
"builtin.module"() ({
  "func.func"() <{function_type = () -> !torch.vtensor<[4],none>, 
                   sym_name = "invalid_tensor"}> ({
    %0 = "torch.vtensor.literal"() <{value = dense<1> : tensor<4xi32>}> 
      : () -> !torch.vtensor<[4],none>
    "func.return"(%0) : (!torch.vtensor<[4],none>) -> ()
  }) : () -> ()
}) : () -> ()
\end{lstlisting}

\textbf{Root cause:} The Torch dialect's tensor type construction does not validate that the element type is a valid tensor element type. The \texttt{none} type causes a null pointer dereference when the verifier attempts to check element type constraints.

\textbf{Expected behavior:} Type construction should validate element types and reject invalid types like \texttt{none} with a diagnostic error.

\subsection{Stack Overflow: Circular SSA Dependencies in Dialect Conversion}

\textbf{Dialect:} Arith (during conversion to LLVM) \\
\textbf{Component:} Dialect conversion framework \\
\textbf{Symptom:} Stack overflow/segmentation fault when processing circular SSA value dependencies during dialect lowering

\textbf{Minimal reproducer (generic format):}
\begin{lstlisting}[language=MLIR, basicstyle=\scriptsize\ttfamily]
"builtin.module"() ({
  %0 = "arith.addi"(%1, %0) <{overflowFlags = #arith.overflow<none>}> 
    : (index, index) -> index
  %1 = "arith.constant"() <{value = 0 : index}> : () -> index
  %2 = "arith.constant"() <{value = 1 : index}> : () -> index
}) : () -> ()
\end{lstlisting}

\textbf{Equivalent assembly format:}
\begin{lstlisting}[language=MLIR, basicstyle=\scriptsize\ttfamily]
module {
  %0 = arith.addi %1, %0 : index
  %1 = arith.constant 0 : index
  %2 = arith.constant 1 : index
}
\end{lstlisting}

\textbf{Root cause:} The program violates SSA form: \texttt{\%0} is used in its own definition (\texttt{arith.addi \%1, \%0}), and \texttt{\%1} is used before being defined. During dialect conversion with \texttt{-convert-ub-to-llvm}, the conversion framework attempts to recursively resolve value mappings, entering infinite recursion when encountering the circular dependency \texttt{\%0 → \%1 → \%0}.

\textbf{Expected behavior:} The MLIR verifier should reject this program during initial verification, catching SSA violations before any dialect conversion begins. If the verifier is bypassed (e.g., with \texttt{--verify-each=false}), the conversion framework should detect circular dependencies and fail gracefully with a diagnostic error rather than stack overflowing.

\textbf{Defense-in-depth failure:} This bug reveals a gap in MLIR's multi-layer validation:
\begin{enumerate}[noitemsep]
\item The parser accepts the syntactically valid but semantically invalid program
\item The verifier (if run) should catch SSA violations but may be bypassed
\item The conversion framework assumes well-formed IR and crashes on malformed input
\end{enumerate}

This demonstrates the importance of robust verification at every stage, as downstream passes should not assume perfect input even within the same compiler infrastructure.
\end{document}